\documentclass[usenatbib]{mn2e}

\newcommand{\sxp}{SXP\,5.05}

\bibpunct{(}{)}{;}{a}{}{,}

\newcommand{\ergs}[1]{$\times 10^{#1}$ erg s$^{-1}$}

\newcommand{\hcm}[1]{$\times 10^{#1}$ cm$^{-2}$}
\newcommand{\ohcm}[1]{$10^{#1}$ cm$^{-2}$}
\newcommand{\expo}[1]{$\times 10^{#1}$}

\newcommand{\ltsima}{$\buildrel < \over \sim$}
\newcommand{\lsim}{\lower.5ex\hbox{\ltsima}}
\newcommand{\gtsima}{$\buildrel > \over \sim$}
\newcommand{\gsim}{\lower.5ex\hbox{\gtsima}}

\usepackage{gensymb}
\usepackage{graphicx}
\usepackage{amssymb}
\usepackage{epsfig}
\usepackage{natbib}
\usepackage{epsf,palatino,url,graphicx,wrapfig,array,setspace,amssymb,fancyhdr,multirow,lscape,appendix,rotating,amsmath,xcolor}

\def\ion#1#2{#1$\;${\small\rm\@Roman{#2}}\relax}

\title[SXP 5.05 ]{SXP 5.05 = IGR J00569-7226 : using X-rays to explore the structure of a Be star's circumstellar disk}

\author[M.J. Coe et al.]{M. J.~Coe$^{1}$, E. S. Bartlett$^{2}$, A.J. Bird$^{1}$, F. Haberl$^{3}$, J. A. Kennea$^{4}$, V.A. McBride$^{2,5}$ \and L.J. Townsend$^{2}$ \& A. Udalski$^{6}$\\
\\
$^{1}$ Physics and Astronomy, University of Southampton, SO17
1BJ, UK. \\
$^{2}$ Astronomy, Gravity and Cosmology Centre, Department of Astronomy, University of Cape Town, Rondebosch, 7701, South Africa. \\
$^{3}$ Max-Planck-Institut f\"ur extraterrestrische Physik,
           Giessenbachstra{\ss}e, 85748 Garching, Germany \\
$^{4}$ Department of Astronomy and Astrophysics, The Pennsylvania State University, University Park, PA 16802, USA. \\
$^{5}$ South African Astronomical Observatory, PO Box 9, Observatory, 7935, South Africa. \\
$^{6}$ Warsaw University Observatory, Aleje Ujazdowskie 4, 00-478 Warsaw, Poland \\
}

\begin{document}

\date{Oct 31 2014}

\pagerange{\pageref{firstpage}--\pageref{lastpage}} \pubyear{2002}

\maketitle

\label{firstpage}

\begin{abstract}

{On MJD 56590-1 (2013 Oct 25-26) observations of the Magellanic Clouds by the {\it INTErnational Gamma-Ray Astrophysics Laboratory (INTEGRAL)} observatory discovered a previously-unreported bright, flaring X-ray source. This source was initially given the identification IGR J00569-7226. Subsequent multi-wavelength observations identified the system as new Be/X-ray binary system in the Small Magellanic Cloud. Follow-up X-ray observations by {\it Swift} and {\it XMM-Newton} revealed an X-ray pulse period of 5.05s and that the system underwent regular occulation/eclipse behaviour every 17d. This is the first reported eclipsing Be/X-ray binary system in the SMC, and only the second such system known to date. Furthermore, the nature of the occultation makes it possible to use the neutron star to ``X-ray'' the circumstellar disk, thereby, for the first time, revealing direct observational evidence for its size and clumpy structure. {\it Swift}  timing measurements allowed for the binary solution to be calculated from the Doppler shifted X-ray pulsations. This solution suggests this is a low eccentricity binary relative to others measured in the SMC. Finally it is interesting to note that the mass determined from this dynamical method for the Be star ($\sim13.0M_\odot$) is significantly different from that inferred from the spectroscopic classification of B0.2Ve ($\sim16.0M_\odot$) - an effect that has been noted for some other high mass X-ray binary (HMXB) systems.
}

\end{abstract}


\section{Introduction and background}

The Be/X-ray systems represent the largest sub-class of all HMXBs.  A survey of the literature reveals that of the $\sim$240 HMXBs known in our Galaxy and the Magellanic Clouds (Liu et al., 2005, 2006), $\ge$50\%
fall within this class of binary.  In fact, in recent years it has emerged that there is a substantial population of HMXBs in the Small Magellanic Cloud (SMC) comparable in number to the Galactic population. Though unlike the Galactic population, all except one of the SMC HMXBs are Be star systems.  In these systems the orbit of the Be star
and the compact object, presumably a neutron star, is generally wide
and eccentric.  X-ray outbursts are normally associated with the
passage of the neutron star close to the circumstellar disk (Okazaki
\& Negueruela 2001), and generally are classified as Types I or II (Stella, White \& Rosner, 1986). The Type I outbursts occur periodically at the time of the periastron passage of the neutron star, whereas Type II outbursts are much more extensive and occur when the circumstellar material expands to fill most, or all of the orbit. General reviews of such HMXB systems may
be found in Reig (2011), Negueruela (1998), Corbet et al. (2009) and Coe et al. (2000, 2009).

One of the aims of the {\it INTEGRAL}
SMC survey  is the ongoing
study of the Be/X-ray binary population of the SMC,
which can be used as a star formation tracer for $\sim$50
Myr old populations (Antoniou et al. 2010). In this paper we
present the analysis of X-ray and optical data from a newly
discovered X-ray source which we catalogue as IGR J00569-7226 (Coe et al., 2013b). This source exhibits most of the characteristics of Be/X-ray binary systems, revealing both a binary period of 17d (Coe et al., 2013c) and an X-ray pulse period of 5.05s (Coe et al., 2013d). Because of this pulse period we will refer to the system as SXP 5.05 throughout this paper (following the naming convention of Coe et al., 2005). Unusually, the neutron star in this system is clearly occulted by the Be star's circumstellar disk. In addition, there is evidence for a complete eclipse of the neutron star by the Be star - if so, this will be the first eclipsing Be/X-ray system in the SMC. The only previously known ``eclipsing'' Be/X-ray binary system is the recently discovered system, LXP 169, in the LMC (Maggi et al, 2013). But that system only exhibits transits by the neutron star across the face of the Be star and no actual eclipses by the Be star of thew neutron star - presumably possible if the orbital eccentricity is large enough.

\begin{figure*}
\includegraphics[angle=90,scale=0.7]{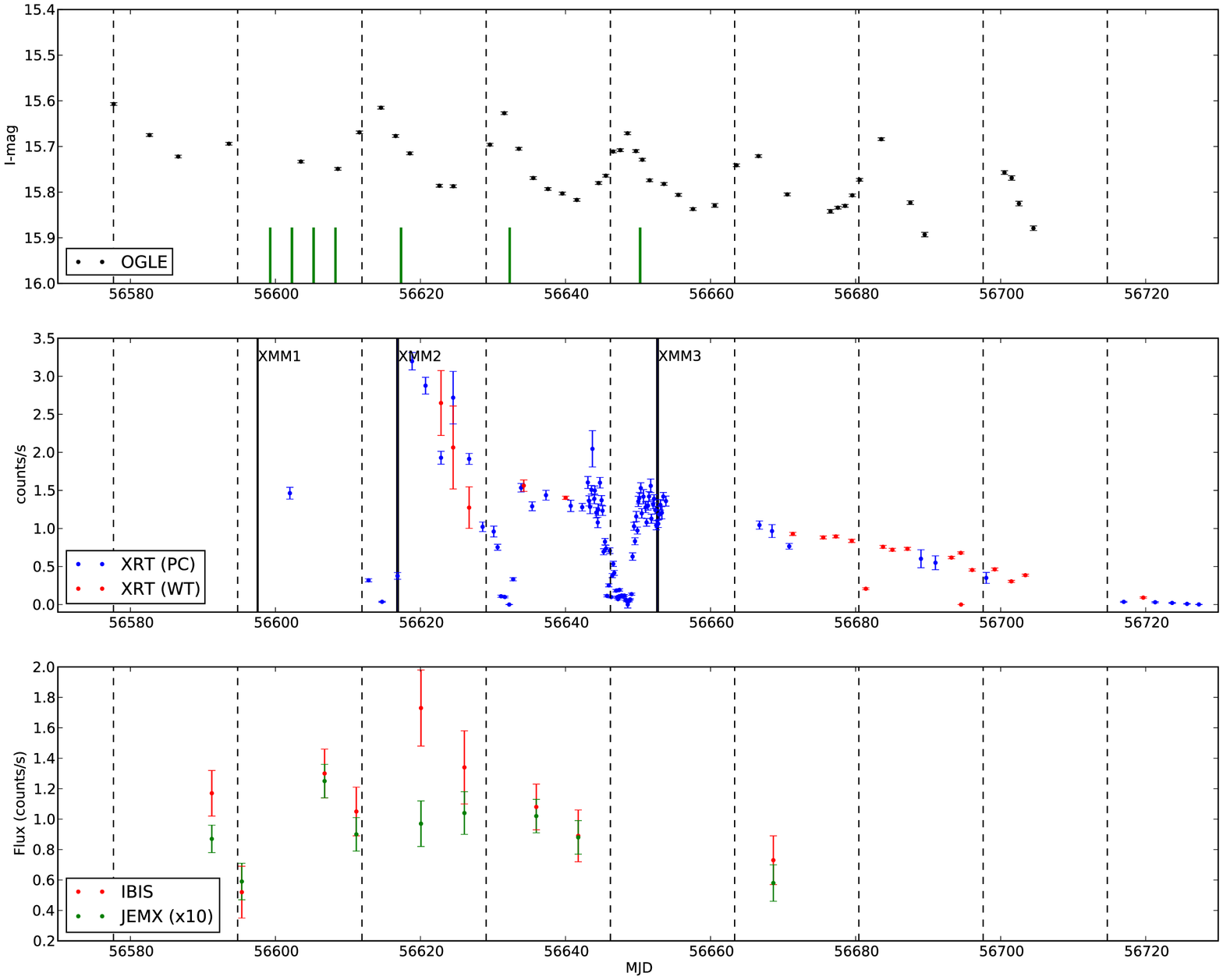}
\caption{Summary of all observations of SXP 5.05. The dashed vertical lines indicate the position of periastron of the neutron star based upon the ephemeris presented in Table 5.
Top panel: OGLE I-band photometry with the timings of the SALT spectroscopic measurements indicated as green vertical bars. Middle panel: {\it Swift}  X-ray flux measurements with the positions of the detailed XMM-Newton measurements indicated by blue vertical lines. Bottom panel: {\it INTEGRAL} hard X-ray measurements from both the IBIS and JEM-X instruments.
}
\label{fig:sxp505_summary}
\end{figure*}

Shown in Figure~\ref{fig:sxp505_summary} is an overall summary of the behaviour of this source at optical and X-ray wavelengths. Each data set is discussed separately in the following sections and possible models presented in Section 4.

\section{X-ray Observations}

\subsection{INTEGRAL}

SXP 5.05 was detected throughout the 2013 SMC monitoring campaign with the {\it INTEGRAL} observatory (Winkler et al. 2003). The Imager on Board the {\it INTEGRAL} satellite (IBIS)  (Ubertini et al. 2003) is optimised for an energy range of 15--200 keV and has a field of view of 29 $\times$ 29 degrees (Full Width Zero Intensity response), and so is well suited to observing large sky areas for point sources. The Joint European X-ray Monitor (JEM-X) telescope provides additional spectral coverage down to 3keV over a smaller field of view. Observations were performed approximately every two {\it INTEGRAL} orbits, and typically consisted of 2 repetitions of the standard 5 $\times$ 5 dither pattern within each of those orbits. Overall the observations spanned the period from revolution 1347 (MJD 56590, 2013 Oct 24) to revolution 1373 (MJD 56668, 2014 Jan 11). Total IBIS exposure was 888ks, with an overall live-time of 620ks in $\sim$450 pointings.

In order to provide rapid source discovery, near real-time data were processed using the {\it INTEGRAL} Offline Science Analysis (OSA, Goldwurm et al. 2003) v9 software and instrument characteristics files. IBIS (17-35 keV) and JEM-X (3-10 keV) maps were constructed for each revolution of data and searched for excesses. The IBIS energy band was chosen to maximise the detection significance of SMC X-1, and hence other SMC accreting X-ray pulsars, which have similar spectral shapes to SMC X-1 in this energy range. The consolidated data were re-processed using the latest OSA v10 software when they became available.

SXP 5.05 was first seen in revolution-level mosaics from {\it INTEGRAL} revolution 1347 (Oct 24th 2013), at a detection significance of 7.6$\sigma$ in the IBIS mosaic and 9.5$\sigma$ in the JEM-X mosaic. An initial position estimate was taken from the combined JEM-X1/JEM-X2 mosaic using the {\textsf j\_src\_locator} tool, and was (RA,dec)=(14.239,-72.431) with a position uncertainty $\sim$2$'$.

In the following observations in revolution 1349 (Oct 30th 2013), SXP 5.05 was observed to have faded significantly in both IBIS and JEM-X, dropping to 3$\sigma$ in the IBIS map (this is below the usual source detection limits) and 4.9$\sigma$ in JEM-X. The changes in the IBIS and JEM-X broad-band fluxes between revolutions 1347 and 1349 were indicative of a change in spectrum of the source.

In revolution 1352 (9 Nov 2013), SXP 5.05 was seen to brighten again, reaching its highest detected flux levels in both 3-10keV and 17-35keV bands, both of which were $\sim$25\% higher than the discovery levels in rev 1347. Thereafter, the source flux was seen to reduce slowly in both IBIS and JEM-X energy bands. By the time of revolution 1373 (Jan 11 2014), the source was once again formally below the IBIS detection levels (at 4.3$\sigma$) but still detected in JEM-X (4.6$\sigma$).
The overall evolution of the fluxes from IBIS and JEM-X (combined JEM-X1 and JEM-X2 maps) are shown in Figure~\ref{fig:sxp505_summary}.

\begin{figure*}
\includegraphics[angle=-90,width=180mm]{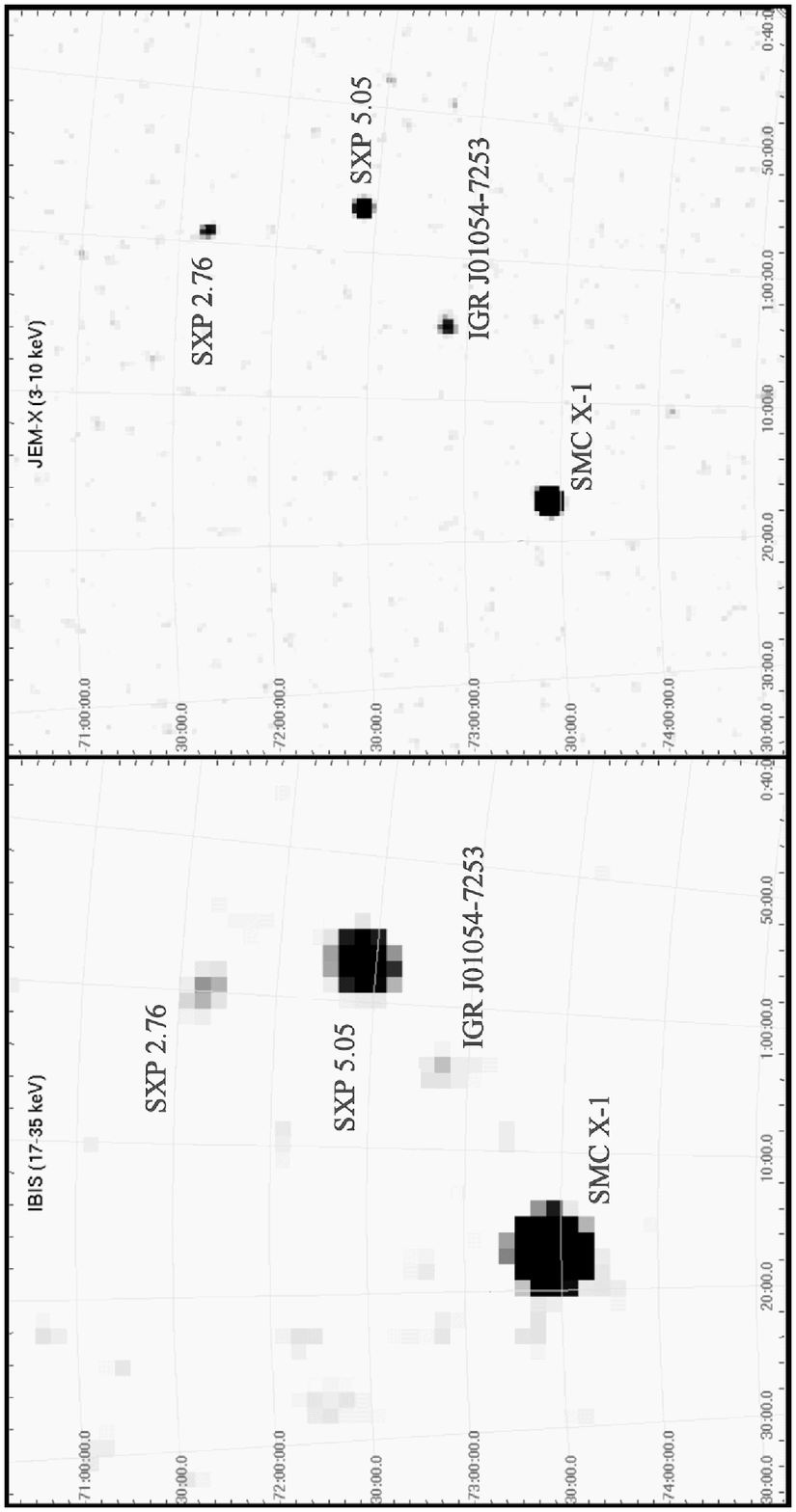}
\caption{Summed {\it INTEGRAL} images from November 2013 - January 2014 observations covering 4 x 4 degrees in size. Left: IBIS 17-35 keV significance image.
Right: combined JEM-X1 + JEM-X2 3-10 keV significance image.
}
\label{fig:int_final}
\end{figure*}

Shown in Figure~\ref{fig:int_final} are the images from the summed total data set of the {\it INTEGRAL} monitoring campaign. From these images four sources are clearly visible and one more lies in the Wing of the SMC:

(a) IGR J00569-7226 (=SXP 5.05) which is the subject of this paper.

(b) SMC X-1 which varied in brightness over the 2-3 months of observing sessions following the familiar super-orbital and binary modulation (Coe et al. 2013a).

(c) SXP 2.76 (= RX J0059.2-7138) is a well-established source originally identified by Hughes (1994) and Southwell \& Charles (1996).

(d) IGR J01054-7253 (=SXP11.5) another well-established source seen in previous {\it INTEGRAL} surveys of the SMC (Coe et al. 2010).

(e) IGR J01217-7257 a bright previously-unknown source seen in just the last observation (11 January 2014) lying in the Wing of the SMC. Follow-up {\it Swift}  observations located this source at a position of
RA(J2000) = $01h 21m 41s$, Dec(J2000) =
$-72^\circ 57' 33''$, with an estimated uncertainty of $4''$ radius
(Coe et al, 2014). This position was coincident with an OGLE IV optical object SMC732.03.3540 which revealed an 84d modulation. No X-ray pulsations were seen.

\subsection{XMM-Newton}

\subsubsection{Observations}\label{xmm:obs}

The \emph{INTEGRAL} monitoring campaign came with three supporting {\it XMM-Newton} Target of Opportunity (ToO) observations. Data from the European Photon Imaging Cameras (EPIC) and the Reflection Grating Spectrometer (RGS) were processed using the XMM-Newton Science Analysis System v13.0.0 (SAS) along with software packages from FTOOLS v6.15.1. Table\begin{table*}
\centering
\caption{\emph{XMM-Newton} EPIC observations of SXP 5.05.}\label{table:obs}
\begin{tabular}{ccccccccc}
\hline\noalign{\smallskip}
ObsID & Camera & Filter & Read Out & \multicolumn{3}{c}{Observation} & Exp.$^{(a)}$ & Count Rate $^{(b)}$\\\noalign{\smallskip}
& & & Mode & Date (MJD) & Start(UT) & End(UT) & (ks) & counts~s$^{-1}$ \\
\noalign{\smallskip}
\hline\noalign{\smallskip}
\multirow{3}{*}{0700580101} & MOS1 & Thin & Full Frame      & \multirow{3}{*}{56597} & 10:39 & 14:24 & 13.5 & \textbf{$0.297\pm0.004$}\\\noalign{\smallskip}
                                              & MOS2 & Thin & Full Frame      &                                     & 10:40 & 14:24 & 13.5 & \textbf{$0.286\pm0.004$}\\\noalign{\smallskip}
                                               & pn       & Thin & Small Window &                                     & 11:53 & 14:27 & 9.2   & \textbf{$0.96\pm0.01$}\\\noalign{\smallskip}
\hline

\multirow{3}{*}{0700580401} & MOS1 & Thin & Full Frame & \multirow{3}{*}{56616} & 17:15 & 22:41 & 19.6  & \textbf{$1.95\pm0.01$}\\\noalign{\smallskip}
                                              & MOS2 & Thin & Full Frame &                                     & 17:15 & 22:41 & 19.6  & \textbf{$1.94\pm0.01$}\\\noalign{\smallskip}
                                              & pn 	  & Thin & Full Frame &				       & 17:37 & 22:49 & 18.7 & \textbf{$7.02\pm0.02$}\\\noalign{\smallskip}

\hline

\multirow{3}{*}{0700580601} & MOS1 & Thin & Full Frame      & \multirow{3}{*}{56652} & 13:49 & 19:00 & 18.6 &  \textbf{$2.71\pm0.01$}\\\noalign{\smallskip}
					    & MOS2 & Thin & Full Frame      &                                     & 13:50 & 19:00 & 18.6 &  \textbf{$2.69\pm0.01$}\\\noalign{\smallskip}
					    &pn        & Thin & Small Window &				   & 13:55 & 19:05  & 18.5 &  \textbf{$11.66\pm0.03$}\\\noalign{\smallskip}
\hline
\end{tabular}\newline
\flushleft{\textbf{Notes.} $^{(a)}$ Before any filtering has been applied. $^{(b)}$ {0.2-10.0~keV}. From source detection, i.e. after filtering has been applied.}
\end{table*} \ref{table:obs} summarises the details of the EPIC MOS (Turner et al. 2001) and pn (Str{\"u}der et al. 2001) observations.

The first observation (referred to as Obs1 hereafter) was performed MJD 56597( 2013 November 1), 7 days after the initial \emph{INTEGRAL} detection. The \emph{INTEGRAL} JEM-X count rate, 
1.6 counts~s$^{-1}$, implied an EPIC-pn 0.2-10.0~keV count rate of $\sim19\text{ counts s}^{-1}$ (assuming a simple absorbed power law model with n$_H=6\times10^{20}\text{ cm}^{-2}\text{ and }\Gamma=1.0$):  above the 2 counts~s$^{-1}$ pile up threshold for the detector in full frame mode\footnote{http://xmm.esac.esa.int/external/xmm\_user\_support/\\documentation/uhb/epicmode.html}. To avoid pile up, the EPIC-pn was used in small window mode which offers superior timing resolution but with a reduction in the field of view (4.4\arcmin$\times$4.3\arcmin as opposed to $\sim30$\arcmin diameter) and ontime of the telescope. The field of view of the EPIC-pn detector is similar to the $1\sigma$ error circle of the \emph{INTEGRAL} source position (see Section 2.1). This presented a serious problem for source localisation. However, we were made aware of a new transient, with position consistent with the \emph{INTEGRAL} error circle, detected in the \emph{Chandra} large survey of the SMC (Haberl, priv comm.). The decision was made to use this position as the telescope pointing.

The count rate of the source in Obs1 was considerably less than expected (see Table \ref{table:obs}). As such, the decision was made to leave the EPIC-pn detector in the default full frame readout mode for the second observation (Obs2), taken MJD=56616 (2013 November 20). The final observation (Obs3) of SXP 5.05 by \emph{XMM-Newton} was made MJD=56652 (2013 December 26). Again, based on the source activity in the period since the Obs2, the EPIC-pn detector was returned to small window mode. Figure \begin{figure*}
\centering
\includegraphics[width=1.0\textwidth]{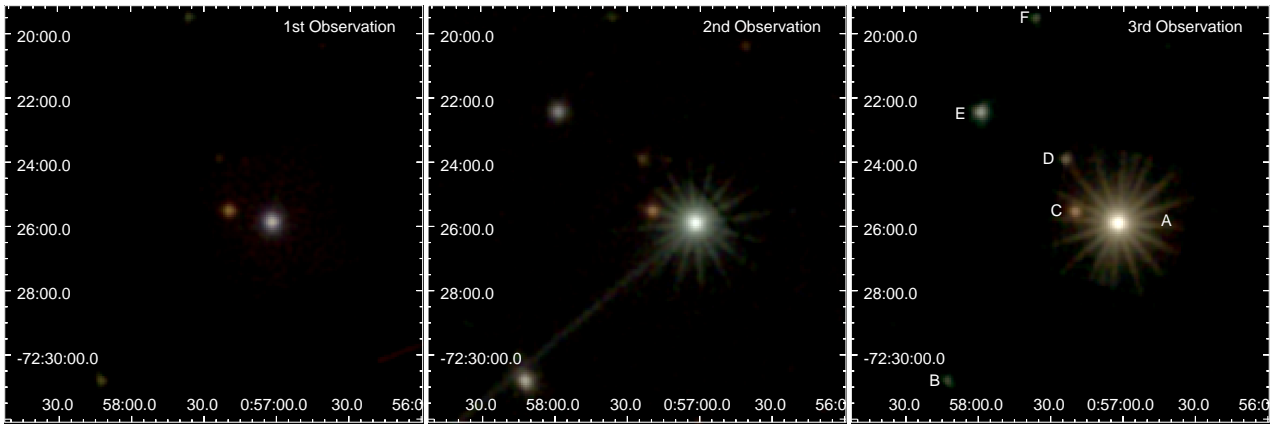}
\caption{False colour mosaic images of the three EPIC detectors. Red corresponds to 0.2-2.0~keV, green to 2.0-4.5~keV and blue to 4.5-10.0~keV. All images have had a Gaussian smoothing applied, with a width of 2 pixels, and have the same colour scale for comparison. Sources identified in the field of view are: \textbf{A.} SXP 5.05 (the subject of this paper), \textbf{B.} SXP 293, \textbf{C.} A known quasar with z=0.15, identified in both the \emph{XMM-Newton} and \emph{Chandra} point source catalogues, \textbf{D.} HMXB candidate XMMU J005724.0-722357, \textbf{E.} SXP 7.92, a recently identified Be/X-ray binary system (Israel et al, 2013), \textbf{F.} SXP 565.
}\label{fig:colour}
\end{figure*}\ref{fig:colour} shows false colour images of the three EPIC observations which reveal the detection of 4 other Be/X-ray systems in the same field as SXP 5.05. One of these objects, SXP 7.92, is recent {\it Chandra} discovery (Israel et al, 2013) and these \emph{XMM-Newton} data will be the subject of a more detailed paper (Israel et al, 2014).

\subsubsection{Data Reduction}\label{xmm:data}

The EPIC-MOS  and pn observational data files were processed with \textsf{emproc} and \textsf{epproc} respectively. The data were screened for periods of high background activity by examining the \textgreater10 keV count rate. No filtering was required for the first and third observation. The first 5.9~ks of Obs2 were affected by high background. These intervals were removed from our analysis. 

Cleaned images were then created in 5 energy bands: 0.2-1.0~keV, 1.0-2.0~keV, 2.0-4.5~keV, 4.5-10.0~keV and 0.2-10.0~keV for each observation.  A box sliding detection was performed simultaneously on all 12 images (4 bands $\times$ 3 detectors) twice (the first with a locally estimated background the second using the background map) with the task \textsf{eboxdetect}, followed by the maximum likelihood fitting using the task \textsf{emldetect}. This process resulted in a list of sources, including their positions, errors and background subtracted counts, for each of the three observations.

As well as searching for sources in each observation individually, we also performed the source detection algorithm on all three observations simultaneously (i.e. 36 images) using the tasks \textsf{emosaic\_prep} and \textsf{emosaicproc}.

The average count rate of Obs2 exceeds the threshold for pile up for the read out mode. The observation was checked for pile up using the SAS task \textsf{epatplot}. Despite the high count rate no evidence for pile up was found.

Light curves and spectra of SXP 5.05 were extracted from the EPIC-pn detector with the same energy ranges as described above. Despite the variations in the source count rate, a extraction radius of 30\arcsec was used in all observations to avoid any contamination from the neighbouring source. ``Single'' and ''double'' pixel events were selected with quality flag (FLAG=0), i.e. all bad pixels and columns were disregarded. Photon arrival times were converted to barycentric dynamical time, centred at the solar system barycenter, using the SAS task \textsf{barycen}.  For the first and third observation, the background region fell on the same CCD chip and had an extraction radius of 30\arcsec (identical to that of the source region). For Obs2, the background region fell on a neighbouring CCD in the pn detector. The extraction radius was 100\arcsec.

The count rate of the source was high enough in the second and third observation to allow RGS spectra to be extracted.  Despite the flare identified at the start of Obs2 in the EPIC instruments, no filtering was required for either instrument for either observation.

\subsubsection{X-ray Position}\label{xmm:posn}

The source detection described in Section \ref{xmm:obs} was performed on all three EPIC cameras for each of the \emph{XMM-Newton} observations, both individually and simultaneously. Table \begin{table}
\centering
\caption{Position of SXP 5.05 as determined by \emph{XMM-Newton} }\label{table:posn}
\begin{tabular}{lccc}
\hline
\multirow{2}{*}{Obs} 	& RA 				& dec 	  							& $\sigma_{stat}$ \\\noalign{\smallskip}
 					& (J2000) 			& (J2000) 							& arcsec \\\noalign{\smallskip}
 \hline\noalign{\smallskip}
 Obs1 				& $00^h57^m2.24^s$ 	& $-72\degree25\arcmin53.7\arcsec$ 		& 0.2 \\\noalign{\smallskip}
 Obs2			        & $00^h57^m2.20^s$	& $-72\degree25\arcmin54.35\arcsec$		& 0.02 \\\noalign{\smallskip}
 Obs3 				& $00^h57^m2.32^s$ 	& $-72\degree25\arcmin54.99\arcsec$ 		& 0.01 \\\noalign{\smallskip}
 \hline\noalign{\smallskip}
 Combined			& $00^h57^m2.30^s$	& $-72\degree25\arcmin54.75\arcsec$		& 0.02 \\
 \hline
 \end{tabular}
 \end{table}\ref{table:posn} lists the position of SXP 5.05 determined in each observation as well as the final position determined from fitting all 45 images simultaneously, along with the 1$\sigma$ statistical error, $\sigma_{stat}$. The 1$\sigma$ systematic uncertainty was assumed to be 1 arcsec in accordance with the findings of the \emph{XMM-Newton} Serendipitous Source catalogue (Watson et al. 2009). This is the dominant component of the error on the positions. These positions are consistent with that reported by Kennea (2013). Figure \begin{figure}
\centering
\includegraphics[width=0.5\textwidth]{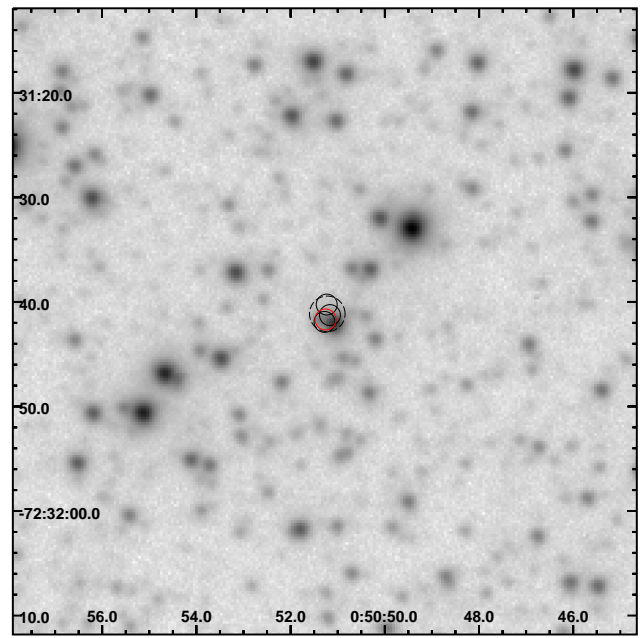}
\caption{A $1\times1$ arcmin \emph{I}-band image of SXP 5.05 taken from the OGLE IV data. The smaller, solid black circles represent the 1$\sigma$ error circles of the three individual \emph{XMM-Newton} ToOs. The red solid circle is the source position as determined by simultaneous source detection on all three observations. The broken circle is the \emph{Swift} 90\% confidence radius.\label{fig:optical}}
\end{figure} \ref{fig:optical} shows a \emph{I}-band image with the location of the \emph{Swift} and \emph{XMM-Newton} positions with radii equal to the 1$\sigma$ errors.

\subsubsection{Timing Analysis}\label{xmm:timing}

\begin{figure*}
\centering
\includegraphics[width=1.0\textwidth]{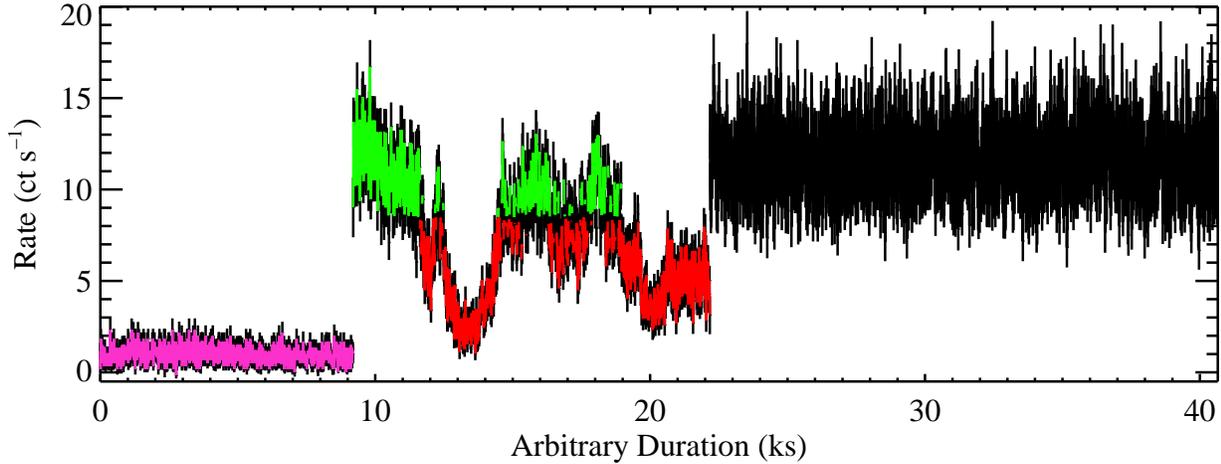}
\caption{The three separate \emph{XMM-Newton} EPIC-pn 0.2-10.0 keV light curves concatenated together. Obs1 is magenta, high count rate intervals of Obs2 are in green, low count rate intervals of Obs2 in red and Obs3, black. The colours correspond to the spectra shown in Section \ref{xmm:spectra}}
\label{fig:xyz}
\end{figure*}

Figure \ref{fig:xyz} shows the EPIC-pn 0.2-10.0~keV light curves of all three observations alongside each other. Each observation is characterised by its own distinctive behaviour (see Table \ref{tab:timing}).

\begin{table}
 \centering
 \caption{Summary of the timing analysis of all 3 \emph{XMM-Newtion} observations. The fractional root-mean-squared value has even calculated according to Equation (10) of Vaughan et al. (2003)}\label{tab:timing}
 \begin{tabular}{lcccc}
 \hline
 \multirow{2}{*}{Obs} 	& Count Rate				 & Fractional	 & Pulse Period	 & Pulsed \\\noalign{\smallskip}
  					& ct s$^{-1}$				 & rms		 & (s)			 & Fraction \\\noalign{\smallskip}
  \hline\noalign{\smallskip}
  Obs1 				& $0.95\pm0.01$ 			& 	0.11		& -				 		& $<$\textbf{0.35}$^{(a)}$\\\noalign{\smallskip}
  Obs2 				& $7.36\pm0.03$			&	0.39		& 5.05066$\pm$0.00003		& \textbf{0.26$\pm$0.03} \\\noalign{\smallskip}
  Obs3 				& $11.60\pm0.03$			& 	0.09		& 5.04843$\pm$0.00001		& \textbf{0.56$\pm$0.06} \\\noalign{\smallskip}
  \hline
 \end{tabular}\newline
 \flushleft{\textbf{Notes.} $^{(a)} 3\sigma$ upper limit.}
 \end{table}

Figure \begin{figure}
\includegraphics[width=0.5\textwidth]{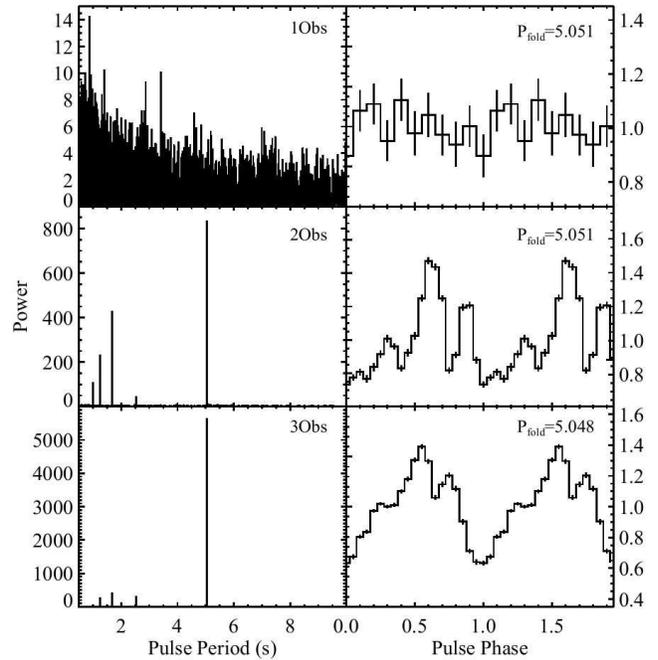}
\caption{Lomb-Scargle periodograms and normalised pulse profiles of each \emph{XMM-Newton} EPIC-pn 0.2-10.0~keV observation in descending order. The profiles have been arbitrarily aligned in phase. Obs1 shows no evidence for any pulsations and so the light curve has been folded on the period detected in Obs2.}\label{fig:pprofiles}
\end{figure}\ref{fig:pprofiles} shows the Lomb-Scargle periodograms of the EPIC-pn 0.2-10.0~keV 0.1~s binned light curves and the normalised pulse profiles of each observation. Unusually for a Be/X-ray, the light curve of Obs1 shows no evidence for pulsations. Pulsations are clearly detected in Obs2 and Obs3 at 5.051~s. and 5.048~s with Lomb-Scargle power 833.4 and 5644.8 respectively.

Monte Carlo simulations were used to determine the errors on the pulse periods: 1000 new light curves were created for both Obs2 and Obs3 by varying each point in the original light curves within their 1$\sigma$ errors using a Gaussian random number generator. Each of these light curves were searched for a period between 0.1 and 10~s in the same manner as the original light curves (i.e. Lomb-Scargle analysis). The histograms of the resulting periods are well fit with Gaussian functions with mean 5.05066 and 5.04843 and standard deviations $3\times10^{-5}$ and $1\times10^{-5}$. Thus we determine the pulse period of SXP 5.05 to be 5.05066$\pm$0.00003~s, as of MJD=56616, and 5.04843$\pm$0.00001~s, as of MJD=56652.

The pulsed fractions of the light curves were calculated by integrating over the pulse profile and taking the ratio of the pulsed component of the profile (i.e. the area of the pulse profile above the profile minimum) and the total count rate. The light curves of Obs2 and Obs3 were split into four energy bands (0.2-1.0~keV, 1.0-2.0~keV, 2.0-4.5~keV and 4.5-10.0~keV) to investigate if there was/is any energy dependence in the shape of the pulse profile or the pulsed fraction. Within each observation, the normalised pulse profiles (i.e. the pulsed profiles divided by the average count rate in the energy bands) above 1.0~keV are consistent within the 1~$\sigma$ errors. The 0.2-1.0~keV pulse profiles are consistent with the $>1.0$~keV pulse profiles within the 2$\sigma$ errors.


Both Obs2 and Obs3 were also investigated to see if there was any evidences for temporal variations in the hardness ratio (and hence the spectra). The hardness ratio is defined as:\begin{equation}
 \frac{C_{hard}-C_{soft}}{C_{hard}+C_{soft}}
\end{equation} where $C_{hard}$ and $C_{soft}$ are the count rates in the 2.0-10.0~keV and 0.2-2.0~keV bands respectively. The hardness ratio can take any value between -1.0 (i.e. all source counts in the 0.2-2.0~keV energy range) and 1.0 (all counts in the 2.0-10.0~keV energy range). No significant variations were found on the timescale of the pulse period, however, Obs2 shows a clear anti-correlation between the total count rate of the source and hardness (see Figure\begin{figure*}
\centering
\includegraphics[width=1.0\textwidth]{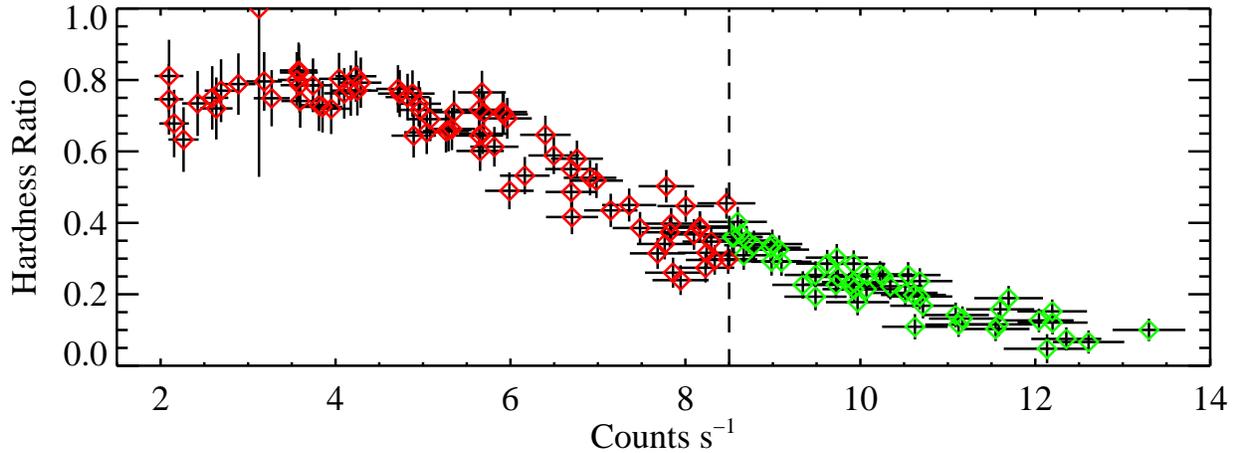}
\caption{Variation of hardness ratio with count rate for Obs2. The broken line marks the count rate at which we split the data into hard and soft intervals}\label{fig:Obs2_rate_hr}
\end{figure*} \ref{fig:Obs2_rate_hr}): The source becomes appreciably harder at lower count rates. The light curve was split into two intervals, a soft/high count rate interval (indicated in green in Figures \ref{fig:xyz} and \ref{fig:Obs2_rate_hr}) a hard/low count rate interval for spectral analysis (see Section \ref{xmm:spectra}).

\subsubsection{Spectral Analysis}\label{xmm:spectra}

To investigate the cause for the spectral changes between the three \emph{XMM-Newton} observations and during the second observation, as indicated by the hardness ratio variations, we simultaneously fitted the EPIC-pn spectra with the same model, only allowing certain model parameters to vary. We included the total spectra from Obs1 and Obs3 and the two spectra from Obs2 (see Section \ref{xmm:timing}), defined from hard/low count rate and soft/high count rate intervals. The spectra from Obs2 and Obs3 are very similar in shape and intensity at energies above 6 keV, indicating that absorption is the main parameter responsible for the spectral changes at lower energies.

\begin{figure*}
  \resizebox{0.75\hsize}{!}{\includegraphics[angle=-90,clip=]{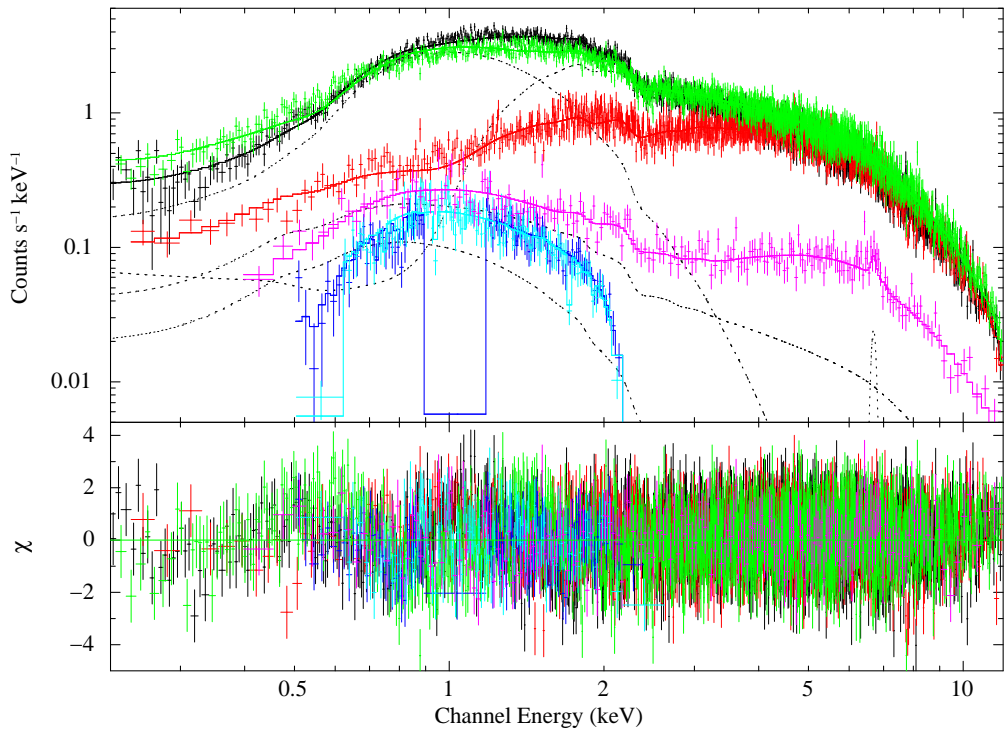}}
  \caption{The X-ray spectra of \sxp\ as observed during Obs1 (magenta), Obs2 (hard/low count rate intervals in red, soft/high count rate intervals in green) and Obs3 (black) as observed by EPIC-pn. The RGS spectra of Obs3 are shown in blue and light-blue. The upper panel presents the spectra and best fit model (as solid line) while the residuals are plotted in the lower panel. The individual model components for Obs3 are indicated by dashed lines.
          }
  \label{fig:xfit}
\end{figure*}

To fit the spectra we started with a model consisting of a power-law and a soft component. The soft component dominates the spectra in the energy band of about 0.5 to 2 keV, depending on the amount of absorption seen in the power-law component. We tested a black-body model for this component, but found unrealistically high values for the size of the emission region to be compatible with an origin at the neutron star surface. Therefore, we used a multi-temperature accretion disk model (\emph{diskbb} in \textsf{XSPEC}) which provides a similar fit quality. At very low energies (below 0.5 keV) another, even softer component is visible in the spectra. In eclipsing super-giant high mass X-ray binaries such soft emission is visible due to scattering and re-processing of the X-ray photons originating from the vicinity of the neutron star in the strong stellar wind, when the direct X-ray emission along the line of sight is strongly suppressed by high
absorption. This component was forced to be at an intensity of 2.5\% of the direct emission components (including both power-law and disk blackbody), however, attenuated by lower absorption than the direct emission. To better constrain the parameters of the \emph{diskbb} component (and to verify that it also represents higher-resolution spectra) we included the RGS spectra from Obs3 in the fit (all model parameters were linked with those for the corresponding EPIC-pn spectrum). No constant factor to allow for calibration uncertainties between the instruments was necessary to fit the RGS spectra. For each of the three emission components we applied absorption with free column densities in the fit with elemental abundances of 0.2 solar (for elements heavier than He) together with an overall absorption (fixed at 6\hcm{20} with solar abundances following Wilms et al. (2000) accounting for the Galactic foreground absorption. The model with its best fit parameters is summarised in Table~\ref{tab:xfit}. The fit
results in a reduced $\chi^2_{\nu}$ of 1.08 for 4461 degrees of freedom. The spectra together with best-fit model are shown in Fig.~\ref{fig:xfit}.

\begin{table*}
  \caption[]{Results of the simultaneous fit to the four EPIC-pn spectra described in the text, Obs1, Obs2 soft interval, Obs2 hard interval and Obs3. All errors are evaluated at the 90\% confidence level. Fluxes and luminosities are given for the 0.2-10 keV band (distance 60 kpc).}
    \begin{tabular}{lcccccccccc}
      \hline\hline\noalign{\smallskip}
      \multicolumn{1}{l}{Obs} &
      \multicolumn{1}{c}{N$_{\rm H, db}$ $^{\rm (a)}$} &
      \multicolumn{1}{c}{kT$_{\rm db}$ $^{\rm (a)}$} &
      \multicolumn{1}{c}{K$_{\rm db}$ $^{\rm (a)}$} &
      \multicolumn{1}{c}{N$_{\rm H, pl}$ $^{\rm (b)}$} &
      \multicolumn{1}{c}{$\Gamma$ $^{\rm (b)}$} &
      \multicolumn{1}{c}{K$_{\rm pl}$ $^{\rm (b)}$} &
      \multicolumn{1}{c}{N$_{\rm H, sc}$ $^{\rm (c)}$} &
      \multicolumn{1}{c}{K$_{\rm sc}$ $^{\rm (c)}$} &
      \multicolumn{1}{c}{F$_{\rm x}$} &
      \multicolumn{1}{c}{L$_{\rm x}$} \\
      \multicolumn{1}{c}{} &
      \multicolumn{1}{c}{\ohcm{22}} &
      \multicolumn{1}{c}{keV} &
      \multicolumn{1}{c}{} &
      \multicolumn{1}{c}{\ohcm{22}} &
      \multicolumn{1}{c}{} &
      \multicolumn{1}{c}{\expo{-3}}&
      \multicolumn{1}{c}{\ohcm{21}} &
      \multicolumn{1}{c}{\%} &
      \multicolumn{1}{c}{erg cm$^{-2}$ s$^{-1}$} &
      \multicolumn{1}{c}{erg s$^{-1}$} \\
       \noalign{\smallskip}\hline\noalign{\smallskip} \vspace{.7mm}
      Obs3      & 0.63$\pm$0.03           & 0.42$\pm$0.03   & $23.9_{-5.8}^{+8.1}$ &  5.7$\pm$0.5     & 1.53$\pm$0.02 &  9.9$\pm$0.3 & $1.7_{-0.4}^{+0.7}$ & 2.5           & 6.0\expo{-11}& \textbf{4.3\expo{37}} \\ \vspace{.7mm}
 Obs2 soft & 0.74$\pm$0.04         &             -   &            -         &  9.9$\pm$0.8     &         -     & 12.3$\pm$0.4 & 0.1$\pm$0.1         & 2.5           & 6.7\expo{-11}& \textbf{5.2\expo{37}} \\ \vspace{.7mm}
      Obs2 hard & 5.8$\pm$0.3           &             -   &             -        & 23.7$\pm$1.6     &         -     & 10.0$\pm$0.4 & 1.4$\pm$0.2         & 2.5           & 4.4\expo{-11}& \textbf{4.4\expo{37}} \\ \vspace{.7mm}
      Obs1      & $28.4_{-3.1}^{+3.7}$ &             -   &            -         & $85_{-12}^{+14}$ &         -     &  1.7$\pm$0.2 & 3.9$\pm$0.4         & 2.5$^{\rm d}$ & 6.8\expo{-12}& \textbf{1.3\expo{37}} \\
      \noalign{\smallskip}\hline
    \end{tabular}

  \flushleft{\textbf{Notes.}Spectral model in \textsf{XSPEC} convention: \emph{phabs}$<$1$>$\emph{(vphabs}$<$2$>$*\emph{diskbb}$<$3$>$ + \emph{vphabs}$<$4$>$*\emph{powerlaw}$<$5$>$ + \emph{vphabs}$<$6$>$*\emph{(powerlaw}$<$7$>$ + \emph{diskbb}$<$8$>$\emph{+guassian}$<$9$>$\emph{))}
  $^{\rm (a)}$ Direct disk blackbody component. Normalisation K$_{\rm db}$ in (R$_{\rm in}$ / D$_{10}$)$^2$ $\times$ cos($\theta$)
     with inner disk radius R$_{\rm in}$ in km, distance D$_{10}$ in units of 10 kpc and $\theta$ the angle of the disk ($\theta$ = 0 is face-on).
  $^{\rm (b)}$ Direct power-law component. Normalisation K$_{\rm pl}$ in [photons keV$^{-1}$ cm$^{-2}$ s$^{-1}$ at 1 keV].
  $^{\rm (c)}$ Scattered component with fixed normalisation relative to the direct component (which includes \emph{powerlaw} and \emph{diskbb}).
  $^{\rm d}$ The scattered component for Obs1 was fixed relative to the direct components of Obs3.}

  \label{tab:xfit}
\end{table*}

The simultaneous spectral modelling of the \emph{XMM-Newton} observations demonstrates that the variability seen in the EPIC-pn light curve of Obs2 can solely be explained by absorption effects. The spectra  obtained from Obs3 and the soft interval of Obs2 are very similar with comparable absorption column densities for the accretion disk and the power-law components. In contrast, the spectrum extracted from  hard interval of Obs2 is strongly attenuated by a column density in excess of \ohcm{23} for the direct power-law component.

Correcting the spectra for their absorption results in similar 0.2-10.0 keV luminosities between 4.3 and 5.2\ergs{37} (assuming a distance of 60 kpc) for the spectra from Obs3, Obs2 hard and Obs2 soft, respectively. The RGS spectra reveal no significant emission lines which would be expected from optically thin gas surrounding the neutron star and re-processing the hard X-ray emission. Hickox et al. (2004) argue that for brighter systems a soft excess in the X-ray spectrum is most likely due to radiation from the inner part of the accretion disk. The fact that the absorption column densities we derive from our spectral modelling is lower for the disk component than for the power-law component suggests the soft emission originates further from the neutron star, consistent with an optically thick accretion disk. From the normalisation of the \emph{diskbb} model we derive an inner disk radius of 29.3 km, assuming an inclination of 90 degrees (i.e. accretion disk and orbital plane aligned and viewed close to edge-on).

The spectrum of Obs1 is much harder in comparison to the expected scattered component seen from the other two observations. A significant bump around 5 keV indicates a relatively strong power-law component with very
high absorption, even higher than that in the hard interval of Obs2. This is confirmed by the combined analysis of the EPIC-pn spectra. Overall, the spectrum of Obs1 is well reproduced by the same model, including the direct
\emph{diskbb} and \emph{powerlaw} components, with extreme column densities (see Table~\ref{tab:xfit}), in addition to the scattered component (assumed to be the same as that seen in the spectrum of  Obs3). The presence of direct emission is inconsistent with a full eclipse of the X-ray source by the Be star, but the extremely high absorption measured from the direct emission components suggests that the line of sight passes the Be star near its rim through the very dense innermost part of the Be disk. At high energies (above ~6 keV) the power-law component is not affected by absorption, but is subject to Thomson scattering which reduces the direct emission (this process is not included in the \emph{phabs} model of \textsf{XSPEC}) significantly at column densities approaching \ohcm{24}. The observed flux between 7.0 and 10.0 keV for Obs1 is 16\% of that seen during Obs3. If the intrinsic source flux is similar during the two observations then an electron column density of 2.7\hcm{24} is required to explain the flux reduction.

Another spectral feature expected from absorption and reprocessing in the gas surrounding the X-ray source is Fe-K$\alpha$ fluorescence. A weak line is seen in the EPIC-pn spectrum of Obs1. The line energy was determined from the fit to 6.69$\pm$0.06 keV, consistent with an origin from highly ionised iron. The equivalent width for Obs1 is 78 eV, while for the other observations it is formally reduced to 10 eV by the strong underlying continuum. In the fit we assume a constant line strength, although no line is required for the spectra of Obs2 and Obs3 as the continuum completely dominates at energies around 6.7 keV. The equivalent width for Obs1 is 78$\pm$40 eV (90\% confidence), while for the other observations it is formally reduced to 12$\pm$7 eV by the strong underlying continuum.

\subsection{{\it Swift} }

\subsubsection{\it{Swift} archival observations}

Including all data where SXP 5.05 was inside the {\it X-ray Telescope (XRT)} (Burrows et al., 2005) field of view
(all observations were pointed within $8'$), {\it Swift}  has serendipitously
observed a region including SXP 5.05 on five occasions before the
beginning of the November 2013 outburst.  We have combined these
observations in order to determine if SXP 5.05 has been previously
detected. These data give a combined exposure of $\sim19$\,ks of {\it Proportional Counter} (PC) mode
data, taken between 2006 November and 2010 September. We find no point
source inside the {\it INTEGRAL} error circle, and calculate a $3\sigma$ upper
limit of $7 \times 10^{-14}\ \mathrm{erg\,s^{-1}\,cm^{-2}}$ ($0.5 -
10$\,keV), constraining the pre-outburst flux to less than $3 \times 10^{34}
\mathrm{erg\,s^{-1}}$, assuming a standard SMC distance of 60\,kpc, or less
than $0.02\%$ Eddington luminosity assuming a $1.4\mathrm{M_\odot}$ neutron
star compact object.

\subsubsection{\it{Swift} detection and localization of SXP 5.05}

The first {\it Swift} post-outburst observation of SXP 5.05 started on MJD 56601 (2013
November 5) at 23:42UT, with a 1 ks PC mode observation. Inside the {\it INTEGRAL}
error circle, we found a single bright point source, and assumed to be the
X-ray counterpart of SXP 5.05.

We determined the position of SXP 5.05 utilizing the methods
described by Goad et al. (2007) and Evans et al. (2009). We find that the coordinates
of SXP 5.05 are RA(J2000) = $00h 57m 02^s.34$, Dec(J2000) =
$-72^\circ 25' 55''.34$, with an estimated uncertainty of $2.6''$ radius
(90\% confidence).

\subsubsection{\it{Swift} Timing Analysis}

For analysis of the pulsar timing of SXP 5.05, we utilize only
data taken using the {\it XRT} ;
Windowed Timing (WT) mode, which provides high time resolution (2.2ms)
event time-tagging. Photon Counting (PC) mode data in this case are not
useful as they suffer both from pile-up and aliasing issues, as the PC mode
frame time (2.506s), is very close to 50\% of the mean pulsar period of SXP 5.05. Exposure times are typically between 3-6 ks each.

All {\it Swift XRT} timing analysis was performed using the HEAsoft 6.14. The
individual downloaded datasets for each observation were first reprocessed
through \texttt{xrtpipeline} with the coordinates for SXP 5.05
given, and then event files were extracted using \texttt{xselect},
utilizing a circular extraction region of 20 pixels radius centered on the
coordinates of SXP 5.05. A barycentric correction was then applied
using \texttt{barycorr}.

Period searching was performed using a $Z^2_2$ search (Buccheri et al, 1983),
with the period search centered on $P=5.05$\,s, a step size of
$10^{-5}$\,s, over 10000 steps, searching a period range of $5.0 - 5.1$\,s.
The reported period for each observation corresponds to the peak of the
$Z^2_2$ distribution. In order to double check the determined period we
also ran the period search with \texttt{efsearch}, which utilizes a
$\chi^2$ folding search. We found the results to be consistent, although
the $Z^2_2$ search proved to be more sensitive to detecting the period
where the total counts in the exposure were low. The detected periodicities
are plotted in Figure~\ref{fig:data}.

In order to estimate errors on the measured period, we implemented a Monte
Carlo approach similar to that described by Gotthelf, Vasisht \& Dotani (1999). Based on the
folded light-curve of the WT event data, we create a simulated event file
with the same total event rate and average event rate as the {\it Swift}  observations.
These are then period searched in the same way as above. This process is
repeated 1000 times, and the resultant scatter in the measurement of the
period in these simulated datasets is used to derive the error on the
measured period. We found that in all cases the errors were larger than the
$10^{-5}$s step size used in the period search, showing that no benefit
would be obtained from searching with a higher time step.

We note that after 2014 March 4, the periodicity could no longer be
detected with sufficient accuracy due to the dropping brightness of SXP 5.05.

The orbital motion of a neutron star about its optical companion presents a window through which to study the orbital parameters of that binary system. In Be/X-ray binaries, the orbital motion is most typically seen through the Doppler shifting of the X-ray light emitted by the pole of the neutron star, as the neutron star travels along its orbit. By decoupling the change in this period caused by the orbital motion from the changes due to the accretion torques exerted on the neutron star, we can model this motion and determine the system parameters. This method has been used extensively in the Milky Way to calculate these parameters for several HMXBs, and more recently in the SMC to find the parameters for the first group of systems outside the Galaxy (Townsend et al., 2011a, 2011b).

The details of the Swift/XRT measured periods are presented in Table~\ref{tab:periods}.

\begin{table*}
\label{tab:periods}
\caption{Journal of {\it Swift/XRT} observations done in Windows Timing mode and the determined pulse period for each observation.}
\begin{tabular}{ccccc}
\hline
Observation Date & Observation Span & Period & Exposure & ObsID \\
(mid-point MJD)   & (days) & (s) & (s) & \\
\hline
56640.01981  &   0.10045 & 5.05006 $\pm$ 0.00004 & 7161 & 00033038021 \\
56671.34285  &   0.34258 & 5.04791 $\pm$ 0.00001 & 3822 & 00033038089 \\
56673.20578  &   0.0736  & 5.04868 $\pm$ 0.00008 & 4032 & 00033038090 \\
56675.53490  &   0.07025 & 5.05037 $\pm$ 0.00006 & 3707 & 00033038091 \\
56677.27352  &   0.0711  & 5.05219 $\pm$ 0.00005 & 3780 & 00033038092 \\
56679.47241  &   0.07288 & 5.05387 $\pm$ 0.00004 & 3758 & 00033038093 \\
56681.27895  &   0.14383 & 5.05207 $\pm$ 0.00155 & 3274 & 00033038094 \\
56683.77892  &   0.16333 & 5.04992 $\pm$ 0.00003 & 3552 & 00033038095 \\
56685.07323  &   0.07008 & 5.04873 $\pm$ 0.00007 & 3408 & 00033038096 \\
56687.14618  &   0.07538 & 5.04776 $\pm$ 0.00004 & 3395 & 00033038097 \\
56689.20835  &   0.20087 & 5.04789 $\pm$ 0.00003 & 3900 & 00033038098 \\
56693.21210  &   0.20254 & 5.05070 $\pm$ 0.00002 & 4918 & 00033038100 \\
56694.50240  &   0.30328 & 5.05192 $\pm$ 0.00002 & 6032 & 00033038102 \\
56696.07214  &   0.34128 & 5.05342 $\pm$ 0.00002 & 4330 & 00033038103 \\
56697.68006  &   0.46938 & 5.05291 $\pm$ 0.00001 & 4738 & 00033038104 \\
56699.18124  &   0.10148 & 5.05196 $\pm$ 0.00009 & 2922 & 00033038105 \\
56701.34264  &   0.20489 & 5.04927 $\pm$ 0.00004 & 4264 & 00033038106 \\
\hline
\end{tabular}
\end{table*}

\subsubsection{Binary solutions}

\begin{figure}
 \includegraphics[width=64mm,angle=90]{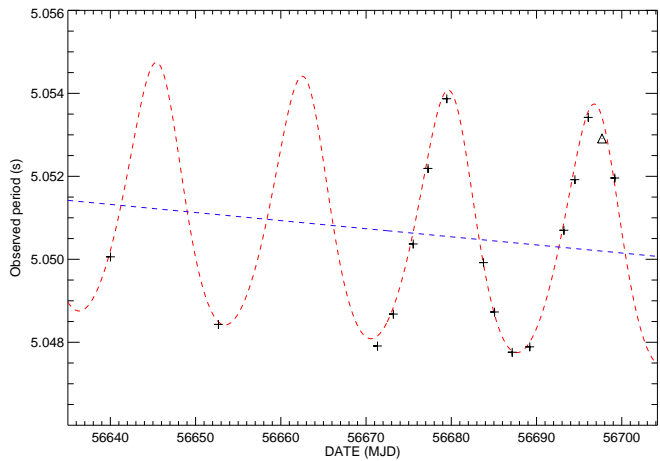}
  \caption{The spin period of SXP 5.05 as measured by \textit{Swift}. Both the orbital motion of the neutron star and the accretion driven spin-up are visible. Overplotted are the spin-up and orbital model from a single fit to the full dataset. It is therefore not the final solution presented in Table 5, but is to be used as an indication of the data used in the fit and the approximate shape and magnitude of the model components. The one point shown as a tiangle was not included in the final fit - see text. We note the independent pulse period determination from {\it XMM-Newton} of 5.0484s on MJD56652 is in complete agreement with the {\it Swift} measurement of the same date.
  \label{fig:data}}
\end{figure}

As the detected spin period will be a function of the orbital parameters, accretion rate and inclination of the system, one needs to be careful in extracting out the required information. In the simplest example of the orbital plane being perpendicular to our line-of-sight (inclination, $i = 0$), any change seen in the spin period of the neutron star should be due to torque produced by accretion onto the neutron star surface. The most common result of this is for the spin period to decrease over the length of the outburst. In the more generic case of the orbital plane being at some angle to our line-of-sight, we see more complicated spin period variations. Convolved with the standard spin-up of the neutron star, there will be a shifting of the X-ray pulse arrival times caused by the orbital motion of the neutron star around the counterpart. It is for this reason that our orbit fitting routine simultaneously fits a simple spin-up model and an orbital radial velocity model.

\begin{eqnarray}
P_{obs} & = & \left(1 + \frac{v(t)}{c} \right) P(t)
  \label{equ:orbit}
\end{eqnarray}

\noindent where $P_{obs}$ is the detected spin period of the pulsar, $c$ is the speed of light and $v(t)$ is the binary radial velocity. $v(t)$ is calculated using the IDL routine {\sc binradvel.pro} from the {\sc aitlib}\footnote{Developed at the Institut f\"{u}r Astronomie und Astrophysik, Abteilung Astronomie, of the University of T\"{u}bingen and at the University of Warwick, UK; see http://astro.uni-tuebingen.de/software/idl/} which uses the procedure of Hilditch et al. (2001) and employs the method of Mikola (1987) to solve Kepler's equation. The spin of the pulsar is given by $P(t)$ which is defined as

\begin{eqnarray}
 P(t) & = & P(t_{0}) + \dot{P}(t - t_{0}) - \frac{\ddot{P}(t - t_{0})^{2}}{2}
  \label{equ:spin}
\end{eqnarray}

\noindent where $\dot{P}$ and $\ddot{P}$ are the spin-up and change in spin-up of the neutron star. We perform a least-squares fit on the data and return the best fitting parameters. The full spin-up component shown in Equation \ref{equ:spin} was fit to the data shown in Figure \ref{fig:data}. One can see from this figure that there is a period between MJD 56670--56700 during which the source was densely monitored. On top of this main data set, there are a few scattered data points over the preceding 30 days. In trying to fit the above model to the full data set, we found that we could not obtain a satisfactory fit. A possible explanation for this is that the accretion rate has changed markedly over the 2 months, as shown in Figure~\ref{fig:sxp505_summary}, meaning there is some higher order period change that we cannot model. The $\ddot{P}$ component can be estimated from the luminosity changes in Figure~\ref{fig:sxp505_summary} and turns out to be roughly $10^{-17} s^2/s$. However, when fitting this component in the model, the parameter always tended to zero. Forcing the parameter to be at this value resulted in even worse fits. To account for this, we decided to only model the data from MJD 56670 (2014 Jan 13) onwards as the effect from the change in flux would be minimised and we would still keep most of the data. We thus removed the $\ddot{P}$ component from the model and only fit a $\dot{P}$ to minimise the number of variables. The result of this was a much more stable fit, with a better $\chi^2$ value. However, the overall $\chi^2_{\nu}$ was still unacceptable, forcing a more thorough investigation into the binary model.

\begin{table}
  \caption{The orbital parameters for SXP 5.05 from the bootstrapping analysis of \textit{Swift} {\it XRT}  data.}
  \label{tab:1}
  \centering
\begin{tabular}{lcl}
  \hline
  Parameter &  & Orbital Solution \\
  \hline
  Orbital period & $\textit{P}_{orbital}$ (d) & $17.13\pm0.14$ \\
  Projected semimajor axis & $\textit{a}_{x}$sin{\it i} (light-s) & $142.4\pm2.5$ \\
  Longitude of periastron & $\omega$ ($^{o}$) & $19.7\pm4.5$ \\
  Eccentricity & $\textit{e}$ & $0.155\pm0.018$ \\
  Orbital epoch & $\tau_{periastron}$ (MJD) & $56680.45\pm0.22$ \\
  First derivative of $\textit{P}$ & $\dot{P}$ ($10^{-10}\mathrm{ss}^{-1}$) & $-2.25\pm0.30$ \\
  \hline
\end{tabular}
\end{table}

\begin{figure}
 \includegraphics[width=64mm,angle=90]{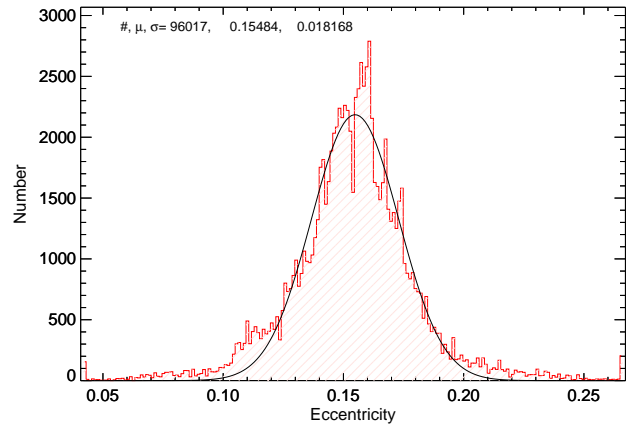}
  \caption{Example distribution from our bootstrapping fitting method. We show the eccentricity parameter, though it is representative of the other parameters. The histogram shows the distribution of the 96,000 fits in the simulation and the fitted Gaussian is overlaid with the mean and error stated.\label{fig:gauss}}
\end{figure}

The reason for a poor fit using the above method seems to result from some random change in the spin period of the neutron star on top of the expected orbital and spin-up signatures. We believe this may arise from a highly variable accretion rate caused by a clumpy wind. Such over-densities in an otherwise uniform wind, would result in unpredictable changes in the accretion rate and, hence, the spin period measured. As this is almost impossible to incorporate into our model, we cannot be sure that the measured period and associated errors can be well described by the model in Equation \ref{equ:orbit}. To account for this, we decided to investigate which data points might be affected by performing a bootstrapping with replacement simulation. The method involved removing a variable number of random points from our dataset and fitting the above model each time to the subset generated, with the expected result being a normal distribution of results for each of the binary parameters. The total number of points in our dataset was 16 and the number of variables 7. The small number of degrees of freedom meant that our distributions were not perfectly Gaussian, but had small bumps and shoulders in some of the parameter distributions. An example distribution and Gaussian fit can be seen in Figure \ref{fig:gauss}. The number of iterations was also kept quite high (100000 runs) to try and mediate this problem. During the initial run, we found one of the 16 data points was causing a poor fit when it was included much more often than the inclusion of the other points. We thus excluded this point (the penultimate one) from the fit and ran the bootstrap again. These results were much more satisfactory and we were able to fit a single Gaussian to all the distributions, to obtain final values and errors. These are presented in Table \ref{tab:1}. The parameter errors are likely quite conservative based on the bootstrapping method used as opposed to a single reduced Chi-squared fit, but rightly so given the difficultly had in fitting these data. The system parameters are discussed further in the latter sections of this paper.

\subsubsection{{\it Swift}  spectral pulse profile information}

\begin{figure}
\includegraphics[angle=-0,width=80mm]{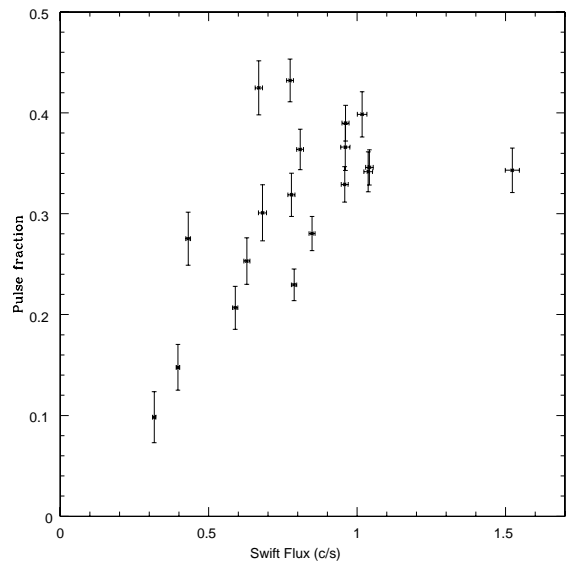}
\caption{Swift/{\it XRT}  pulse fraction versus 0.5--10 keV X-ray count rate.}
\label{fig:pf}
\end{figure}

Shown in Figure~\ref{fig:pf} is the pulse fraction defined as in Equation 1 as a function of the general source brightness.  This figure reveals that the typical pulse fraction in the {\it Swift}  0.5--10 keV energy band is in the 20--40\% range, with strong evidence for a correlation between the pulse fraction and the luminosity.

\begin{figure}
\includegraphics[angle=-0,width=80mm]{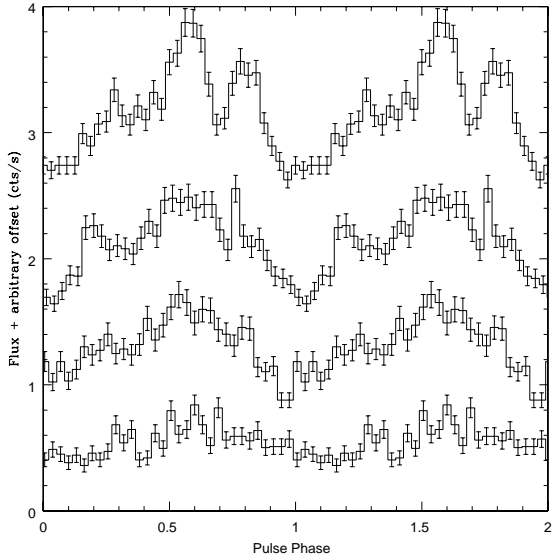}
\caption{Swift/{\it XRT}  pulse profiles (0.5--10 keV). From the top, the profiles are arranged in decreasing source brightness and show observations made on MJD 56639.9, MJD 56671.0, MJD 56683.6 \& MJD 566701.1; the profiles have been separated by adding 2.0, 1.2, 0.5 \& 0 cts/s respectively to their count rates. In addition, the profiles have simply been lined up in phase by eye.}
\label{fig:pshape}
\end{figure}

Figure~\ref{fig:pshape} reveals more details of these pulse profile changes with luminosity. In this figure the decreasing complexity of the pulse profile is revealed as the source moved from a bright state ($\sim$1.5 counts/s) to a fainter state ($\sim$0.5 counts/s). Similar structural changes, but with better signal-to-noise ratios, were reported for EXO 2030+375 by Parmar, White \& Stella (1989). Those authors attributed the changing pulse profiles to be related to a switch in the dominant emission profile from a fan-beam to a pencil-beam as the luminosity changed. Specifically they observed a lack of sharp features in the pulse profile at high luminosity in EXO 20230+375, with the profiles becoming more complex as the luminosity fell from $\sim1\times10^{38}$ erg/s to $\sim1\times10^{37}$ erg/s. Below this flux level the profiles smooth out again, though ultimately the diminishing signal-to-noise ratio as the flux approaches $\sim1\times10^{36}$ erg/s makes this somewhat uncertain.

The brightest {\it Swift}  profile shown in Figure~\ref{fig:pshape} corresponds to a source luminosity of $\sim2\times10^{37}$ erg/s and reveals a strong pulse and inter-pulse peak separated by $\sim$0.2 in phase. This flux level corresponds to that of maximum pulse shape complexity in EXO 2030+375 and hence may support the idea of a combination of pencil and fan beam emissions at this stage. Similar profiles were seen in the {\it XMM} observations - see Figure~\ref{fig:pprofiles}. More detailed modelling, beyond the scope of this paper, including the possibility of an off-set dipole field as suggested by Parmar, White \& Stella (1989) for EXO 2030+375, is required to understand the pulse profile shapes seen here in SXP 5.05.

\subsection{\it RXTE}

The SMC was observed on an almost-weekly basis using the {\it Rossi X-ray Timing Explorer (RXTE)}. The results of searching through these $\sim$13 years worth of  observations of the SMC reveals that SXP 5.05 was frequently within the field of view, often at a high collimator response, but was only detected 5 times at a confidence rating of $\ge$99\% (see Galache et al, 2008 for an explanation of the {\it RXTE} data analysis techniques). The dates detected and the associated periods are presented in Table~\ref{RXTElog}. Using the orbital parameters presented in Table~\ref{tab:1} reveals no obvious phase pattern for the detections, though the uncertainty of 0.14d on the orbital period  means that extrapolating back over a decade makes the phase value increasingly uncertain. Typical errors on a 5-10s pulsar detected by {\it RXTE} at the 99\% confidence level are $\le$0.01s (Galache et al, 2008). The scatter in pulse periods around 5.02-5.09s suggests that the {\it RXTE} data are revealing a pulsar close to equilibrium spin period outside of the times of major outbursts. This would be consistent with the overall behaviour of SMC Be/X-ray binaries seen in the SMC (Klus et al, 2014).

\begin{table}
  \caption{Historical RXTE detections of SXP 5.05 at $\ge$99\% confidence.  }
  \label{RXTElog}
  \begin{tabular}{ccc}
  \hline
  Date (MJD)& Collimator& Period \\
  & Response & detected (s)  \\
  \hline
  55179.3	&0.89	&5.02 \\
  52458.6	&0.26	&5.06 \\
  52882.6	&0.78	&5.07 \\
  53487.9	&0.78	&5.09 \\
  55785.7	&0.89	&5.06 \\

  \hline

  \hline
  \end{tabular}
  \end{table}

\section{Optical observations}
\subsection{OGLE}

Optical Gravitational Lensing Experiment data (OGLE-II, III and IV, Udalski et al. 1997; Udalski 2003) were used to investigate the long-term behaviour of SXP 5.05. OGLE has been regularly monitoring this object since 1997 with the 1.3-m Warsaw telescope at Las Campanas Observatory, Chile, equipped with three generations of CCD camera: a single 2k $\times$ 2k chip operated in driftscan mode (OGLE-II), an eight chip 64 Mpixel mosaic (OGLE-III) and finally a 32-chip 256 Mpixel mosaic (OGLE-IV). Observations were collected in the standard I-band.

OGLE photometric I-band observations of SXP 5.05 were obtained from MJD 50627(1997 June) up to MJD 56689 (2014 January). The OGLE source identification in each phase is given in Table~\ref{tabOgle}.

\begin{table}
  \caption{OGLE identifications for the optical counterpart to SXP 5.05}
  \label{tabOgle}
  \begin{tabular}{ccc}
  \hline
  OGLE project& I-band ID& V-band ID \\
  phase&& \\
  \hline
  OGLE II & smc sc7.266703& - \\
  OGLE III & smc 108.8.26225& smc108.8.v.30808 \\
  OGLE IV & smc 719.18.378& smc719.18.v.340\\
  \hline

  \hline
  \end{tabular}
  \end{table}

\begin{figure}
\includegraphics[angle=-0,width=80mm]{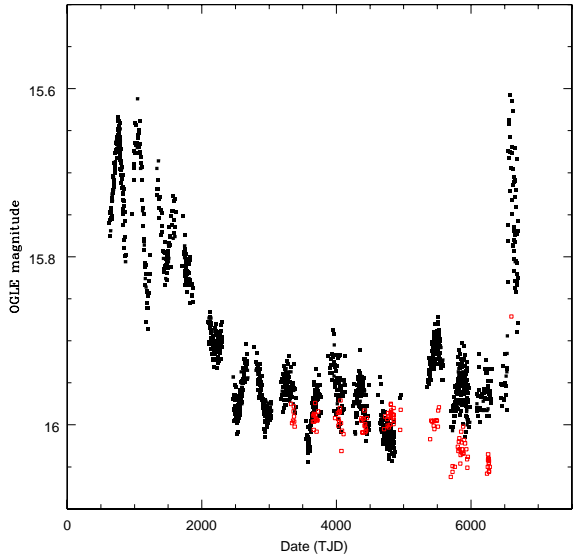}
\caption{OGLE monitoring of SXP 5.05 during the period 1997-2014. The solid black symbols are I-band data, whereas the open red symbols are V-band data.TJD = JD - 2440000.5.
}
\label{fig:ogle_all}
\end{figure}

\begin{figure}
\includegraphics[angle=-0,width=80mm]{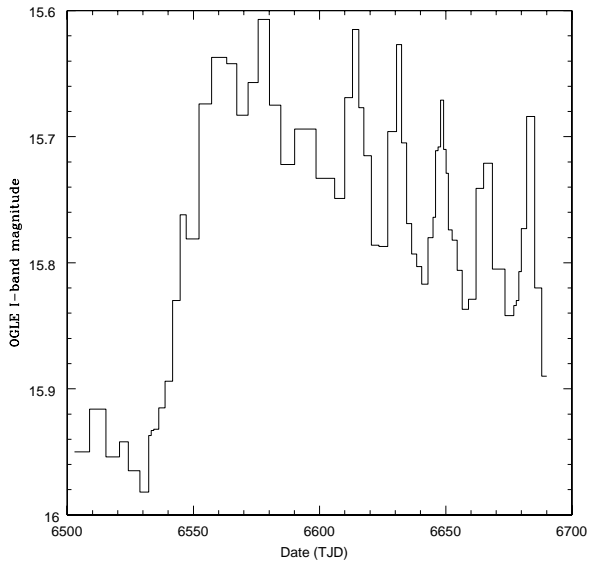}
\caption{OGLE monitoring of SXP 5.05 during the 2013-2014 outburst. The $\sim$17d modulation is very clear.TJD = JD - 2440000.5.
}
\label{fig:ogle_final}
\end{figure}

The full photometry from this 16-17 year coverage is shown in Figure~\ref{fig:ogle_all}. From this figure it is immediately apparent that the source exhibits considerable rapid optical variability on timescales of months to years, culminating in the recent rise of 0.3 magnitudes within a month. Not obvious in this figure is the $\sim$17d modulation seen by close examination of the decline from outburst which occurred at the start and end of this whole data set. This was initially commented upon by Schmidtke \& Cowley (2013) and is shown very clearly in Figure~\ref{fig:ogle_final} during the current outburst.

\begin{figure}
\includegraphics[angle=-0,width=80mm]{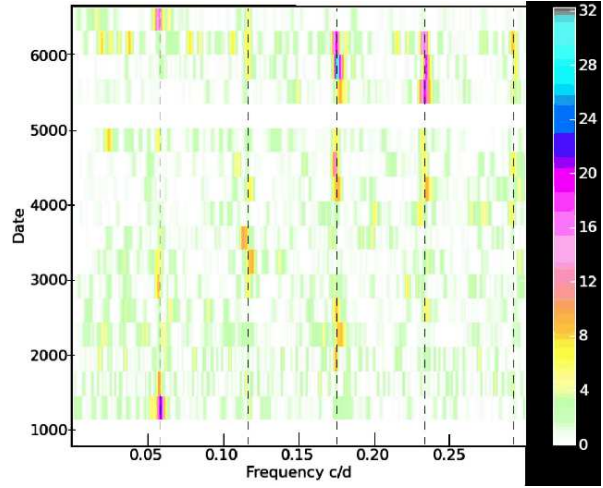}
\caption{Period searching though OGLE data - the fundamental is at a frequency of 0.058 c/d and is indicated by the left-most dashed vertical line. The other vertical dashed lines show the position of various harmonics of the fundamental. The colour scale represents the Lomb-Scargle peak power value.
}
\label{fig:ogle_tony}
\end{figure}

Shown in Figure~\ref{fig:ogle_tony} are the trailed Lomb-Scargle power spectra obtained from each annual sample of OGLE data. The date were detrended and processed using the techniques described in Bird et al. (2012). It is immediately apparent that not only is there power at the fundamental frequency at the start and end of the data run, but in between the main power peaks are often at one of the harmonics. Adding together all the different peaks from all the years gives a final value of $17.13\pm0.01d$ for the fundamental optical period.

\begin{figure}
\includegraphics[angle=-0,width=80mm]{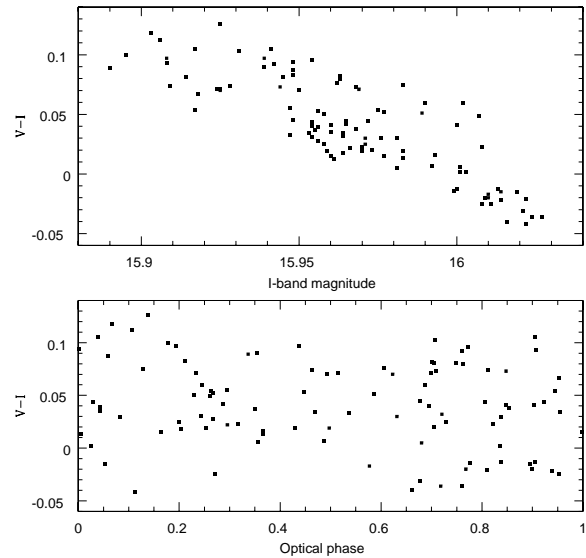}
\caption{OGLE observed colours versus brightness (top panel), and colours versus optical phase (lower panel).
}
\label{fig:ogle_col}
\end{figure}

Shown in Figure~\ref{fig:ogle_col} are the observed (V-I) colours versus the I-band brightness. It is clear there is an obvious correlation between these two parameters as seen in other similar systems (e.g. Vasilopoulos et al. 2014, Coe et al. 2012). The V band data were taken whilst the source was not in X-ray outburst and so this figure shows the general fluctuations in the circumstellar disk that occur most of the time. To check whether the colour variations have anything to do with the neutron star orbit, the colour data are plotted versus the 17.13d X-ray ephemeris in the lower part of Figure~\ref{fig:ogle_col}. There appears to be no relationship with this optical phase and therefore one can conclude that in the quiescent state the neutron star is not exerting any influence over any disk parameters that could result in changing temperatures. However, given the uncertainties on the orbital period quoted in Table 1 there is considerable uncertainty in the phases of the early OGLE data - but even selecting just the most recent few years does not change the lack of pattern in Figure~\ref{fig:ogle_col}.

This is in contrast to that seen in LXP 8.04 where a strong colour effect was seen correlated with the optical phase (Vasilopoulos et al. 2014). Unfortunately there are not enough V band observations during the outbursts of SXP 5.05 just to select those epochs for a more detailed study of the larger disk behaviour.

\subsection{SALT}

SXP 5.05 was observed with the Robert Stobie Spectrograph (RSS) (Burgh et al. 2003, Kobulnicky et al. 2003) on the Southern African Large Telescope (SALT) at various epochs during outburst with a view to spectral classification of the donor star and H$\alpha$ monitoring of the Be star circumstellar disk.  An observation log is presented in Table~\ref{SALTlog}.

\begin{table*}
  \caption{Journal of {\it SALT} spectroscopic observations of SXP 5.05}
  \label{SALTlog}
  \begin{tabular}{cccccc}
  \hline
  Date (MJD)& Grating & Exposure (s) & Resolution (\AA) & Coverage & H$\alpha$ EW (\AA)\\
  \hline
  56598 & PG2300 & 1000 & 2.0 &3800--4900\,\AA\AA & --\\
  56598	& PG1800 & 840 & 4.7 & 5950--7250\,\AA\AA & -6.6 (0.5) \\
  56601 & PG1800 & 840 & 4.7& 5950--7250\,\AA\AA & -7.3 (0.5)\\
  56604 & PG1800 & 840 & 4.7& 5950--7250\AA\AA & -6.6 (0.5)\\
  56607 & PG2300 & 2$\times$1220 & 2.0 & 3800--4900\,\AA\AA& --\\
  56607 & PG1800 & 840 &4.7 & 5950--7250\,\AA\AA & -5.6 (0.5) \\
  56616 & PG1800 & 840 & 4.7 & 5950--7250\,\AA\AA & -8.0 (0.5)\\
  56631 & PG2300 & 1000 & 2.0 & 3800-4900\,\AA\AA & --\\
  56631 & PG1800 & 840 & 4.7 &5950--7250\AA\AA & -6.8 (0.5)\\
  56649 & PG1800 & 1000 & 4.7 & 5950--7250\AA\AA & -7.3 (0.5)\\
  \hline
  \end{tabular}
  \end{table*}

The spectra were corrected for bias and cross talk by the SALT data reduction pipeline (Crawford et al. 2010).  Subsequent data reduction was performed in IRAF\footnote{IRAF is distributed by the National Optical Astronomy Observatory, which is operated by the Association of Universities for Research in Astronomy (AURA) under cooperative agreement with the National Science Foundation}. The two-dimensional images were wavelength calibrated using observations of Neon (PG1800) or Copper Argon arc lamps (PG2300) and then rectified before one dimensional, background subtracted spectra were extracted.  Grating PG2300 spectra were flatfielded.

\begin{figure*}
\includegraphics[width=120mm, angle=90]{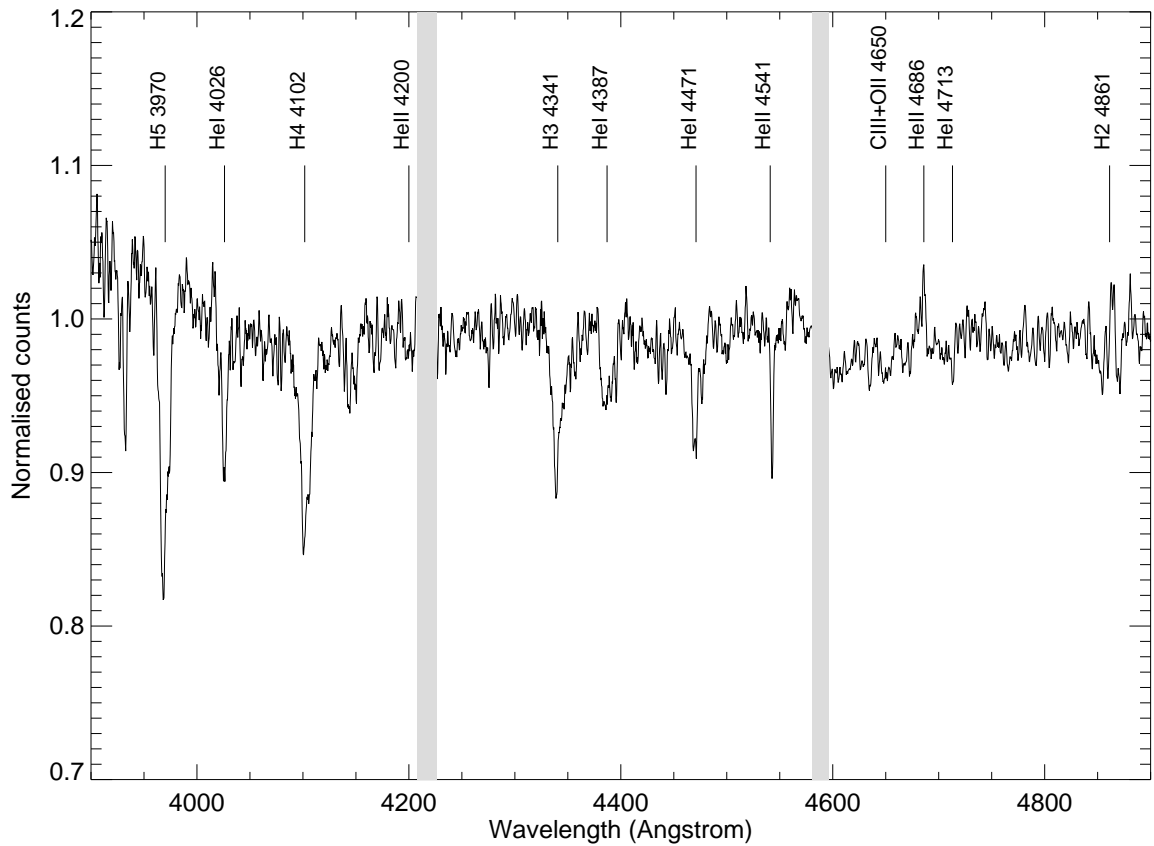}
\caption{The optical spectrum of SXP 5.05, with lines of interest overplotted. The grey shaded areas are the chip gaps in the detector.}
\label{fig:SBS}
\end{figure*}

Figure~\ref{fig:SBS} shows the continuum-normalised, smoothed SALT spectrum observed on 2013-11-02 with grating PG2300, blue shifted by the recession velocity of the SMC (150\,km\,s$^{-1}$).  The Balmer lines and He I lines characteristic of early type stars are clearly visible.  He II at 4541 and 4200 are detected, while H$\beta$ is in emission, which places the spectral type at B0e--B0.2e, according to the classification criteria presented in Evans et al. (2004).  SXP 5.05, which appears to be a member of NGC 330, has been previously observed as part of the VLT-FLAMES survey of massive stars (Evans et al. 2006).  The spectral type determined by those authors is B0.5e.  It is hard to reconcile the SALT data with this slightly later type, but we do note that our result implies a significant discrepancy between the mass determined from the spectral type ($\sim16.0M_\odot$) and the dynamical mass estimates ($\sim13.0M_\odot$) discussed in Section 4.1 below. This system is the first one in the SMC to have a mass determination from both spectral type and dynamical methods and it reveals a substantial gap between the two methods.

The spectrum at that time (between 2003 July and 2004 January) shows very clearly the He II 4686 in absorption consistent with spectral types as cool as B0.5.  Subsequent spectra observed with SALT on 2013-11-11 and 2013-12-05 confirm the He II 4686 line in emission, and we interpret this as a feature arising from an accretion disk around the neutron star.

\begin{figure}
\includegraphics[angle=-0,width=80mm]{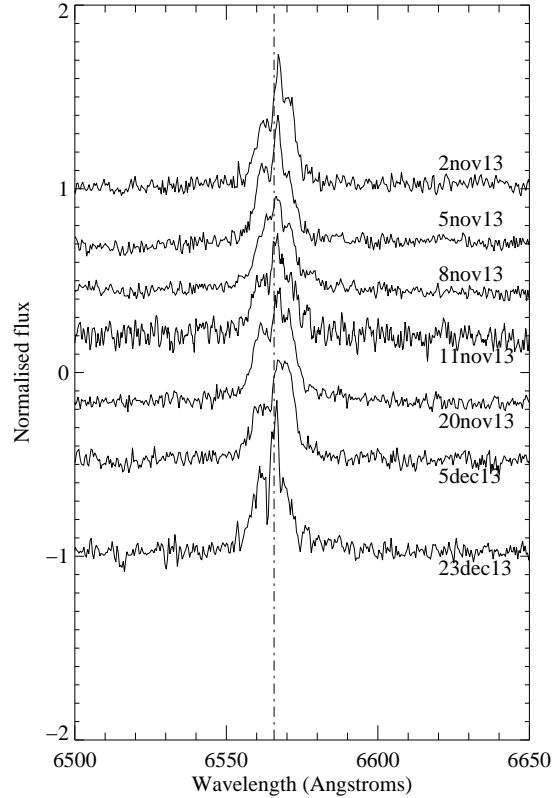}
\caption{H$\alpha$ 6562.3\,\AA \,\,emission line profiles.  Spectra have been rectified with arbitrary offsets on the y-axis to show the line profile evolution with time.
}
\label{fig:salt:ha}
\end{figure}

Shown in Figure~\ref{fig:salt:ha} are the H$\alpha$ profiles. The profiles start off with a clear triple peak structure, then evolves to a double peak structure by
MJD 56631 (2013December 5), but the triple peak structure seems to return by the last observation MJD 56649 (2013 Dec 23). 3 Gaussian lines were fitted to all observations.  The central component is very narrow (sigma $\sim$1\AA) and located at the H$\alpha$ rest wavelength for the SMC, while the two outer
components are broader.  It is worth noting that the VLT-Flames spectra from Evans et al. (2006) also show this triple structure in the H$\alpha$ emission line.

The equivalent width is statistically consistent with a mean value of  $-6.9\pm$0.2\AA ~throughout the observations. Because the SALT coverage is relatively sparse it is not possible to say with any certainty if there is any correlation between the H$\alpha$ EW values and the phase of the OGLE I-band modulation. However, it is perhaps worth noting that the 4 largest H$\alpha$ EW values correspond to peaks in the I-band modulation, and the 3 weakest to the baseline flux level. This suggests that the interaction of the neutron star with the circumstellar disk at periastron creates some disturbance of the disk structure.

The double-peaked H$\alpha$ profile can be used to calculate the radius of the circumstellar disk (Coe et al., 2006).  Using the average peak separation of $8\pm2$\AA, and making the assumptions that the circumstellar disk lies in our line of sight ($i=90^\circ$), and the true mass of the star is $13.0M_\odot$.  The radius is given by:

\begin{equation}
R_\mathrm{circ} =\frac{GM\sin ^2 i}{(0.5\Delta V)^2}
\end{equation}

where $\Delta V$ is the peak separation in velocity space.
This results in a radius of $(2.6\pm1.8)\times10^{10}$\,m  (clearly it can be smaller if the inclination is different).

Another approach to estimating the size of the circumstellar disk is simply to use the H$\alpha$ EW. Assuming the mean value of -6.9$\pm$0.2\AA, the size of the B0V star, and utilising the relationship in Zamanov et al. (2001), the radius of the disk is found to be $(4.8\pm0.1)\times10^{10}$\,m - just about in agreement with the previous number.

\section{Discussion and Conclusions}

\subsection{Binary models}

The binary mass function is described by the following

\begin{eqnarray}
  f_{X}(M) & = & \frac{{4\pi}^{2}a_{x}^{3}\sin^{3}i}{GP^{2}} = \frac{M_{c}^{3}\sin^{3}i}{(M_{x} + M_{c})^{2}}
 \label{equ:xmass}
\end{eqnarray}

\noindent where $M_{x}$ and $M_{c}$ are the masses of the neutron star and the optical counterpart, $i$ is the inclination of the orbital plane to the line of sight, $P$ is the orbital period, $a_{x}$ is the semi-major axis of the ellipse travelled by the neutron star and $e$ is the eccentricity. Thus, if the mass functions of both stars are known, along with the inclination angle, one can dynamically calculate the masses of the two stars. Substituting in the values from Section 2.3.4, we obtain a mass function of 10.56\,M$_{\odot}$. This is quite high relative to many other Be/X-ray binary systems, but not as large as most supergiant systems (see Townsend et al. (2011) for a summary of known binary parameters and mass functions).

It can be seen in Figure~\ref{fig:eclipse_fit} that the time during the eclipse when the flux drops close to zero is at MJD 56648.6. Given the 17.13\,d orbital period found in the OGLE data this point is directly behind the companion star as viewed from Earth. This observation supports our theory that the eclipse (at least the deepest part of the occultation) is caused by the companion. If this is correct, the inclination of this system is almost edge on to our line-of-sight. Taking an orbital phase of 0.04 as the point of eclipse, the separation of the neutron star and the Be star may be determined to be $4.08\times10^{10}$\,m. Combining this with a stellar radius of B0.2V star of $5.2\times10^{9}$\,m gives a minimum orbital inclination of $i=82.7^{\circ}$.

Constraining the inclination in this way can give us an idea of the actual orbital size, the orbit disk mis-allignment, the actual geometry of the binary system and, possibly, the mass of the two bodies. This latter point is very interesting as it would be the first time a neutron star mass measurement will have been dynamically determined in a Be/X-ray binary - however it will require future radial velocity observations of the optical star to determine this unambiguously. Even without this radial velocity curve, we can place constraints on the masses of the two bodies, assuming $i=90^{\circ}$. Given a set of realistic to extreme masses for the neutron star, say, 0.5, 1.0, 1.4, 2.0 and 2.5\,M$_{\odot}$, we can solve the binary mass function for $M_{c}$. The resulting solutions are  $M_{c}$ = 11.5, 12.3, 13.0, 13.8 and 14.5\,$M_{\odot}$ using the respective neutron star masses above. Reducing the inclination to $i=80^{\circ}$ only changes these values by $\le5\%$. Thus, even with an extremely large neutron star mass of 2--2.5\,M$_{\odot}$, the mass of the counterpart is 2--3\,M$_{\odot}$ less than standard mass of a B0.2V star which is around 16\,M$_{\odot}$. Discrepancies of this kind between stellar classification and observed luminosities plus inferred masses \& temperatures, have previously been reported for HMXB companions by Conti (1978) and Kaper (2001). This can only be explored accurately with eclipsing systems, so SXP 5.05 will be a valuable addition to the small number known of such systems.

Assuming an inclination of $i=90^{\circ}$, the actual size of the orbit is given in Table \ref{tab:1}, and is $4.27\times10^{7}$\,km. In Section 3.2 the size of the circumstellar disk was estimated to be 2.6--4.8 $\times10^{7}$\,km, which is similar to the orbital radius and confirms that the neutron star is in contact with the disk during this outburst. This is discussed further below.

The optical flare seen regularly in SXP 5.05 is at a phase of 0.15 relative to periastron passage according to the orbital model. Though delays between optical flares and periastron passages are well established in Be/X-ray systems, this is much later than any other examples we have in the SMC. If we use the X-ray ephemerides from the sources observed as part of the 12 years of observations of the SMC using RXTE (Galache et al, 2008) and compare these to the optical ephemerides determined by Bird et al. (2012) we can derive both optical and X-ray epochs of maximum amplitude and compared them. These results may then be converted into a phase delay for that system based on the optical orbital period - see Table~\ref{tab:delays}. Clearly SXP 5.05 exhibits a much greater phase delay between X-ray \& optical outburst than any of these systems. If we translate the phase delays given in Table~\ref{tab:delays} into elapsed time, then values in the range 2-13d are determined. These values of a few days may well be indicative of some physical characteristic of the circumstellar disk response times within the accretion process.

In addition, we note that the phase of the optical outbursts coincides with that of the X-ray eclipse. We have no explanation for this alignment except to invoke coincidence.

\begin{table}
  \caption{Phase delays between optical outbursts and the X-ray outburst for a small sample of SXP systems. }
  \label{tab:delays}
  \begin{tabular}{ccc}
  \hline
  Source& Phase delay & Orbital period(d) \\
  \hline
SXP 7.78 & 0.04 & 44.8 \\
SXP 46.6 & 0.03--0.10 & 137.4 \\
SXP 91.1 & 0.04 & 88.0\\
SXP 756 & 0.00--0.01  & 394.0\\

  \hline

  \hline
  \end{tabular}
  \end{table}

\subsection{Modelling the circumstellar disk}

\begin{figure}
\includegraphics[angle=-90,width=80mm]{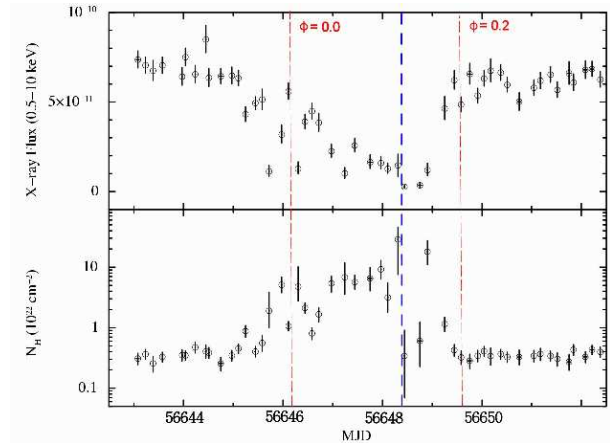}
\caption{Variations in the Swift/{\it XRT}  measured X-ray flux and absorption during the occultation/eclipse. Note that the upper panel has a linear vertical scale, whereas the lower panel is logarithmic. The vertical red dashed lines show the binary phase, and the vertical dashed blue line indicates the binary phase (not the date) of the first {\it XMM-Newton} observation..
}
\label{fig:eclipse_fit}
\end{figure}

Shown in Figure~\ref{fig:eclipse_fit} is a detailed study of the {\it Swift/XRT} flux during the occultation event for which there are the most observations. From this plot it can be seen that the suppression of the source flux occurred gradually over $\sim$120 hours, but in a somewhat erratic manner. After the point of minimum flux, there was a more rapid recovery phase. The total size of the occultation event greatly exceeds the sinmple effect of eclispsing the neutron star with its companion Be star.

 The asymmetry in the X-ray absorption is hard to explain with a circular circumstellar disk in the plane of the neutron star orbit, instead we invoke the possibility of the disk being in a plane orthogonal to the plane of the neutron star orbit. Then, depending upon the orientation of the disk, it is possible to create asymmetries in the absorbing column to the neutron star.

\begin{figure}
\includegraphics[angle=-0,width=80mm]{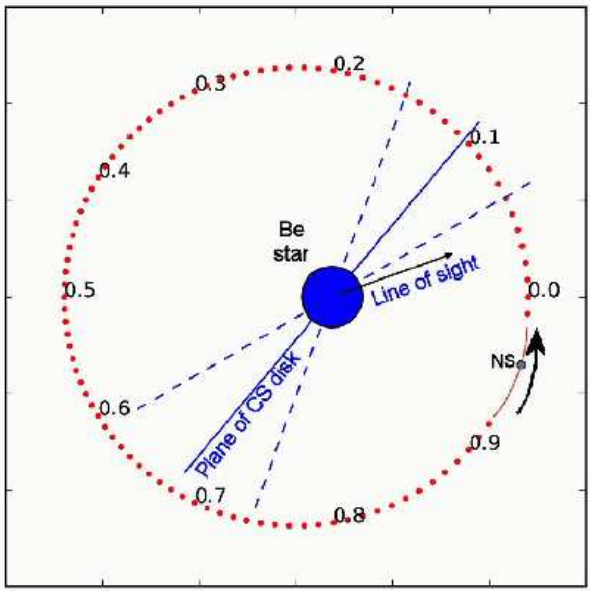}
\caption{The orientation of SXP 5.05 system with respect to our line of sight. Phase 0.0 represents periastron. The plane of the circumstellar disk is orthogonal to the neutron star orbital plane. The blue dashed lines represent the outline of the flared circumstellar disk used in modelling the X-ray data. The solid black arrow represents our line of sight with respect to periastron($\theta_{los}=20\degree$). The dotted black line ($\theta_{disk}=50\degree$) represents the central axis of the circumstellar disk which is inclined orthogonally to the plane of the orbit.
}
\label{fig:mod1}
\end{figure}

\begin{figure}
\includegraphics[angle=90,scale=0.35]{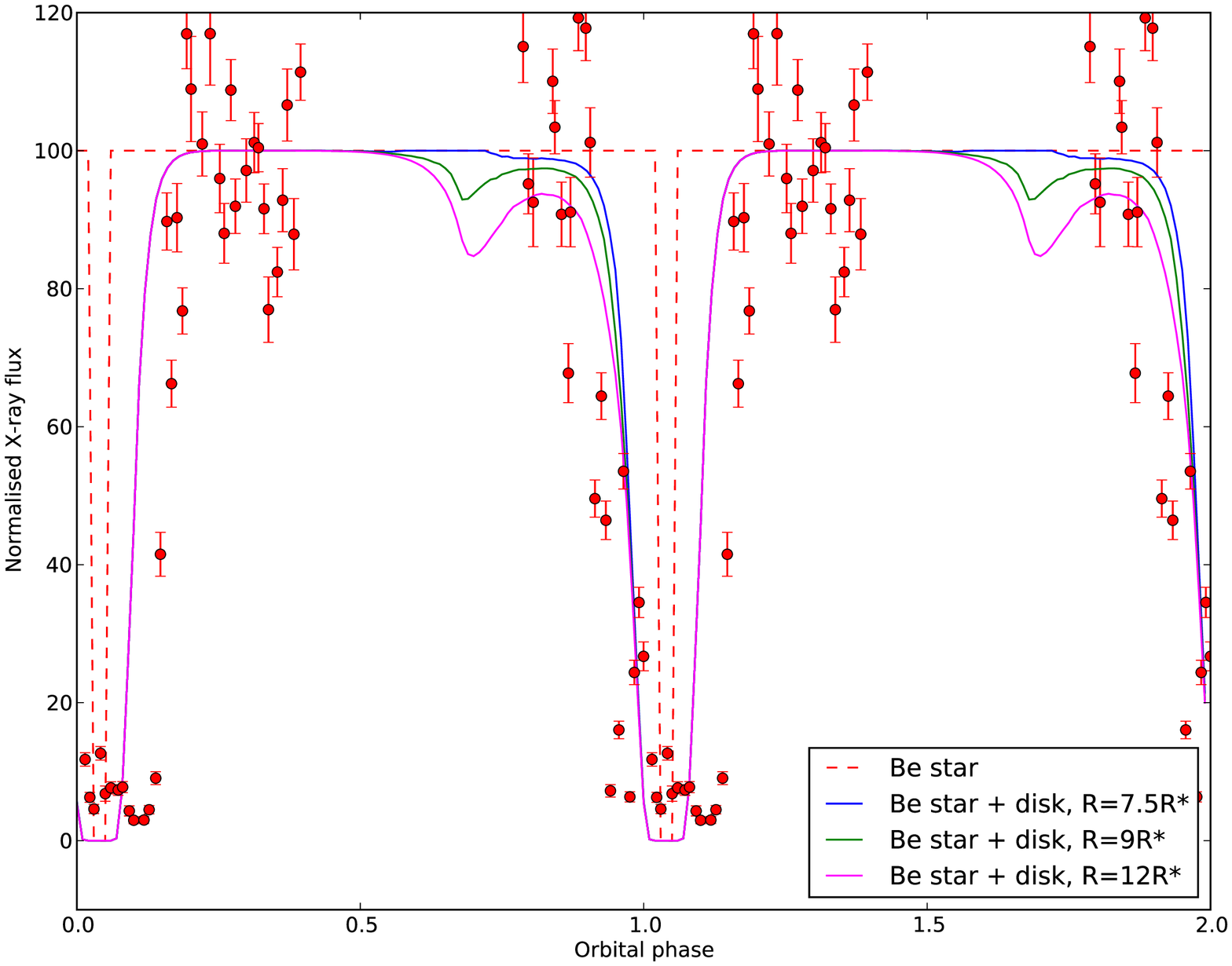}
\caption{Implications of the current model for different disk sizes compared to the {\it Swift/XRT}  data. The red dashed line shows the effect of a simple geometric eclipse by the Be star.
}
\label{fig:mod2}
\end{figure}

We have attempted to model the absorption profile between phases 0.7 and 0.25 by considering the attenuation of the neutron star emission by the Be star's circumstellar disk. Attenuation by the disk is essential to produce the broad orbital effect seen, which is incompatible with an eclipse by the Be star alone.  We require a disk that is partially orthogonal to the orbital plane of the binary, such that the neutron star can move 'behind' the disk with respect to our line of sight for a significant phase range. In order to induce the observed asymmetry in the absorption, we further require that the line of sight is close to the plane of the circumstellar disk, so that the neutron star is observed obliquely through the disk at all times. We have implemented a 2D disk model that is able to capture the key geometric features needed - a truncated, flared disk with a radial density fall-off . We assume the power-law radial profile and simplified disk scale height model from Rivinius, Carciofi \& Martayan (2013), with a linear growth in scale height reaching 3.5R* at 20R*. We implement a truncated circumstellar disk by imposing a hard density drop to zero at the desired distance from the Be star.

With reference to Figure ~\ref{fig:mod1} we can see the neutron star will be obscured by the disk between phases of ~0.7 and~0.2, assuming the black arrow represents our line of sight ($\theta_{los}=50\degree$) and the dotted line ($\theta_{disk}=70\degree$) represents the central axis of the circumstellar disk. These values are only approximate, as the disk scale height prevents any meaningful discussion of the 'thickness' of the disk, and the flaring shown is only for illustration. Any eclipse by the Be star, if present, would be at phase 0.04. For each phase of the neutron star orbit, the cumulative disk density along the line of sight between the neutron star and observer is calculated, and then converted into an attenuation value. Initially (at phase ~0.7) the obscuration is low as the neutron star is only obscured by the low-density outer regions of the flared disk. But then the obscuration increases slowly as the line of sight is through the full thickness of the outer disk. The attenuation then rises rapidly towards phase 0 as the line of sight moves towards the dense central regions of the disk. A much sharper recovery is seen between phase 0.1 and 0.2, as the neutron star traverses the disk again.

From Figure~\ref{fig:sxp505_summary} it can be seen that this occultation event occurred against a general background of the X-ray flux declining globally in a very linear manner between MJD 56635 and the end of the observations. Apart from the occultation event the flux can be accurately described by the simple empirical relationship over this time interval - see Equation~\ref{eq:flux}.

\begin{equation}
\label{eq:flux}
F(t) = -0.01667t + 945.59
\end{equation}

where F(t) is the {\it Swift} /{\it XRT}  count rate, and t is the date in MJD.

Figure ~\ref{fig:mod2} shows a family of curves for a power-law radial profile with an index of n=3.75 and a base density of $\rho_{zero} = 10^{-10} g/cc$ (within the suggested range of Rivinius, Carciofi \& Martayan (2013), but definitely towards the high end, implying that this is a dense disk). The extremes of the curves have been normalised to fit the {\it Swift} /{\it XRT}  count rates scaled according to Equation~\ref{eq:flux}. Three different disk truncation radii are simulated, and the effect on the early stages of obscuration (phase ~0.7) are obvious. More extended disks also produce higher attenuation around phase 0.0, as the line of sight is looking through the disk, although the low density of the outer disk makes this a small effect.

Overall, we conclude that even a simple geometric model of this type is able to explain some of the key features seen in the folded x-ray light curve, principally the long duration over which obscuration is seen, and the asymmetry of the obscuration around the predicted eclipse point. However,we emphasise that this is in no way a parameterised fit to the observed data. Furthermore it is clear that some parts of the model fit do not describe the data very well - specifically the egress phase. Many of the parameters are quite degenerate (notably the disk density and the radial density profile) and we can only state that the 'typical' disk parameters given by Rivinius, Carciofi \& Martayan (2013) do result in a reasonable attenuation profile.

Perhaps the biggest difficulty faced by this simple model is explaining in detail the obscuration observed between phases 0.9 and 1.0. Here we observe flux variations by a factor of 2 in very short timescales which obviously cannot be reproduced by the smooth disk model used here. The presence of such rapid changes in attenuation implies the disk is not homogenous but actually highly structured. We can place limits on the size of these spatial structures by considering the projection of the line of sight through the disk axis. Consecutive measurements at phases 0.964 and 0.975 yielded count rate variations of a factor of 8, these two measurements correspond to lines of sight passing through the disk axis $1.7\times10^{9}m$ (or about 3\% of the circumstellar disk radius) apart. We can therefore conclude that significant density fluctuations are present in the disk at radii of many R*. In particular, we note that there are only large variability in flux/absorption when the NS moves away from us, not when it moves towards us. This raises the possibility that the NS may be producing the inhomogeneities in the circumstellar disk while the it is intrinsically smoother beforehand. Similar effects are proposed for supergiant systems  which also exhibit rapid changes in the absorption column before eclipses (Haberl, White \& Kallman (1989) and  Blondin et al. (1990)).

Finally, by comparing the long term decline in the X-ray flux with that of the optical, both after MJD 56580, it is possible to compare the two levels of flux change. From Equation~\ref{eq:flux} and Figure~\ref{fig:ogle_final} it can be seen that the X-ray flux falls off at a significantly faster rate than the optical flux. In fact, assuming a baseline optical flux for no circumstellar disk of I=16.1, the optical falls by $\sim$13\% over 100d compared to $\sim$67\% for the X-ray. Since the X-ray flux is relying completely upon accreting material from the outer edges of the circumstellar disk, it is to be expected that a shrinking disk will, at some point, no longer be able to feed the accretion process whilst still retaining significant I-band emission. So a modest change in the disk size (and hence I-band emission) produces dramatic changes in the X-ray flux.

\section{Conclusions}

This paper reports the discovery of the first eclipsing Be/X-ray system in the Magellanic Clouds. Because of the eclipsing nature, an accurate orbital model solution has been possible. In addition, detailed observations have been obtained showing the neutron star being occulted by not only the companion Be star, but also some extended structure. we have interpreted this extended structure as the circumstellar disk around the Be star and have presented one possible simple model for how this might be configured. Clearly further exploration of all the orbital parameters combined with a physically more realistic model of the circumstellar disk is needed, but beyond the scope of this initial paper.

\section{Acknowledgements}

The OGLE project has received funding from the European Research Council under the European Community's Seventh Framework Programme (FP7/2007-2013)/ERC grant agreement no. 246678 to AU. VAM acknowledges funding from the NRF, South Africa. LJT and ESB are Claude Leon Foundation Research Fellows. Some of the observations reported in this paper were obtained with the Southern African Large Telescope (SALT).”

\bsp

\label{lastpage}

\end{document}